\documentclass[times,twocolumn,final,authoryear]{tmp}

\usepackage{ycviu}
\usepackage{framed,multirow}

\usepackage{amssymb}
\usepackage{latexsym}

\usepackage{url}
\usepackage{xcolor}
\usepackage{subfig}
\definecolor{newcolor}{rgb}{.8,.349,.1}

\journal{}

\begin{document}

\thispagestyle{empty}

\ifpreprint
  \setcounter{page}{1}
\else
  \setcounter{page}{1}
\fi

\begin{frontmatter}

\title{A Generalized Deep Learning Framework for Whole-Slide Image Segmentation and Analysis}

\author[1,3]{Mahendra \snm{Khened}} 
\author[1,3]{Avinash \snm{Kori}}
\author[1,3]{Haran \snm{Rajkumar}}
\author[2]{Balaji \snm{Srinivasan}}
\author[1]{Ganapathy \snm{Krishnamurthi}\corref{cor1}}
\cortext[cor1]{Corresponding author:}
\ead{gankrish@iitm.ac.in}
\address[1]{Department of Engineering Design, Indian Institute of Technology Madras, 600036 India}
\address[2]{Department of Mechanical Engineering, Indian Institute of Technology Madras, 600036 India}
\address[3]{Authors contributed equally}

\begin{abstract}
Histopathology tissue analysis is considered the gold standard in cancer diagnosis and prognosis. Whole slide imaging, i.e., the scanning and digitization of entire histology slides, are now being adopted across the world in pathology labs. Trained histopathologists can provide an accurate diagnosis of biopsy specimens based on whole slide images (WSI). However, given the large size of these images and the increase in the number of potential cancer cases, an automated solution as an aid to histopathologists is highly desirable. In the recent past, deep learning-based techniques, namely, CNNs, have provided state of the art results in a wide variety of image analysis tasks, including analysis of digitized slides. However, the size of images and variability in histopathology tasks makes it a challenge to develop an integrated framework for histopathology image analysis. We propose a deep learning-based framework for histopathology tissue analysis. We demonstrate the generalizability of our framework, including training and inference, on several open-source datasets, which include CAMELYON (breast cancer metastases), DigestPath (colon cancer), and PAIP (liver cancer) datasets. Our segmentation pipeline is an ensemble of DenseNet-121, Inception-ResNet-V2, and DeeplabV3Plus, where all the networks for each task were trained end to end. Our framework provides segmentation maps given a WSI. Our entire framework and related documentation are freely available at GitHub and PyPi.
\end{abstract}


\end{frontmatter}

\section{Introduction}
\label{path_sec:1}
{H}{istopathology} is considered the gold standard for cancer diagnosis \citep{gurcan2009histopathological,salamat2010robbins} and identification of prognostic and therapeutic targets. Early diagnosis of cancer significantly increases the probability of survival \citep{hawkes2019cancer}. Unfortunately, pathological analysis is an arduous process that is difficult, time-consuming, and requires in-depth knowledge. A study \citep{elmore2015diagnostic} examining breast biopsies concordance among pathologists found that pathologists disagreed with each other on a diagnosis 24.7\% of the time on average. This high rate of misdiagnosis stresses the need for the development of computer-aided methods to aid the pathologists in histopathology.

Digital pathology is the method of digitizing the histology slides to produce high-resolution images \citep{janowczyk2016deep}. Studies have been conducted on the collection, analysis and interpretation of digitized pathological slide images \citep{gurcan2009histopathological}. The increasing prevalence of whole slide imaging (WSI) technology that can scan the entire tissue slide at the subcellular level is making for conducting pathology analysis more viable \citep{madabhushi2016image}. Digital pathology's array of image analysis activities include identification and counting (e.g. mitotic events), segmentation (e.g. nuclei), and tissue differentiation (e.g. cancerous vs. non-cancerous) \citep{janowczyk2016deep,nanthagopal2013classification,guray2006benign}. Segmentation analysis helps in detecting and separating tumour cells from the normal cells \citep{wahlby2004combining,xu2016deep}. Segmentation of whole slide images is usually the precursor for performing various other downstream analyses such as classification and tumour burden estimation.

Automated whole slide image analysis is plagued by a myriad of challenges \citep{tizhoosh2018artificial}, namely:
\begin{enumerate}
    \item Large dimensionality of whole slide images: A WSI digitizes a glass slide at a very high resolution of order 0.25 micrometres/pixel (which corresponds to 40X magnification on a microscope). A typical glass-slide of size 20mm x 15mm results in gigapixel image of size 80,000 x 60,000 pixels.
    
    \item Insufficient training samples: The main impediments to development and clinical implementation of deep learning algorithms consist of sufficiently large, curated, and representative training data which includes expert labelling which is a costly and time-consuming process (e.g., pathologist annotated data). Most clinical research groups currently have restricted access to data. The data is often based on small sample sizes with limited geographic variety which results in algorithms with limited utility and poor generalization.
    
    \item Stain variability across laboratories: As the data is acquired from multiple sources, there exists a lack of uniformity in staining protocol. Building a generalized framework that is invariant to stain pattern variability is challenging.
    
    \item Extraction of clinically relevant features and information: Another major challenge is trying to extract features that are meaningful from a clinical point of view. Deep learning does an excellent task of automatic feature extraction, but understanding these extracted features and extracting meaningful information from them is challenging.
\end{enumerate}

In this paper, a framework developed for analyzing whole slide images of three different cancer sites is presented. The organization of the paper is as follows. Prior work on histopathology image analysis using deep learning-based methods are discussed in section \ref{path_sec:path_prior}. In section \ref{path_sec:path_data} the datasets used in this study are presented.  Discussion on training and inference pipelines is provided in section \ref{path_sec:train_pipeline} and \ref{path_sec:inference_pipeline} respectively. Experimental analysis and comprehensive results are presented in section \ref{path_sec:expi_analysis} and \ref{path_sec:path_results}. Discussion of the results and conclusion of the proposed study with the possible course of research is provided in section \ref{path_sec:discussion}.

\section{Related work}
\label{path_sec:path_prior}
\subsection{Deep learning methods for histopathological image analysis}

The advent of whole slide imaging scanners has enabled digitization of glass-slides at very high resolution. Typical whole slide images are in the  order of gigapixels and usually stored in multi-resolution pyramidal format. These slide images are suitable for developing computer-aided diagnosis systems for automating the pathologist workflow and also with the availability of a large amount of data makes them amenable for analysis with machine learning algorithms. 

In tumour pathology, nuclear morphology and cellular anatomy are often significant determinants of disease severity. In order to make the diagnostic and grading task of tumour less subjective, quantifiable features are derived from the images that correlate with the condition of the disease \citep{gurcan2009histopathological}. For example, algorithms can be designed to detect invasive tumours by first segmenting nuclei from the background, quantifying a number of nuclear characteristics, such as size, shape and distribution, and comparing these characteristics with those of normal cells \citep{diamond2004use}. \citep{yu2016predicting} predicted non-small cell lung cancer prognosis by applying classical machine learning algorithms that use engineered features derived from pathology images.

Feature engineered algorithms rely on a predetermined set of features to classify the tissue and can only classify the tissue as better as the features that differentiate between them, and thus there is a limit to their efficiency, even when there is a large amount of data available to refine the algorithm. Hence, there has been a significant shift in recent years towards the application of deep learning algorithms as they are known for its inherent ability to automatically derive features from its input data. Typical deep learning-based approaches for whole slide image segmentation or classification are usually made by cropping the slide image into multiple small image patches and treating them as independent of each other during training and inference. Furthermore, to make an overall slide-level prediction or to generate a heatmap of regions of interest, patch-level predictions are aggregated in a suitable manner. \citep{cruz2014automatic} was one of the earlier works using this method that showed promising results in detecting invasive ductal breast carcinoma. Several studies have applied deep learning algorithms for various pathology tasks related to breast cancer, prostate biopsies, colon cancer, etc. 

Colorectal carcinoma is the third most common cancer in the world \citep{fleming2012colorectal}. Majority of colorectal carcinoma are adenocarcinomas originating from epithelial cells \citep{hamilton2000carcinoma}. \citep{shapcott2019deep} discuss the application of deep learning methods for cell identification on TCGA data. \citep{kather2019predicting} discuss the deep learning methods to predict the clinical course of colorectal cancer patients based on histopathological images. \citep{bychkov2018deep} discuss the use of Long short-term memory (LSTM) \citep{greff2016lstm} artificial recurrent neural network (RNN) architecture for estimating the patient risk score using spatial sequential memory.

A review on WSI application for histopathological analysis of liver diseases and for understanding liver biology is given by \citep{melo2019whole}. They explore how WSI can enhance the evaluation and quantification of several histologic hepatic parameters and help to identify various liver diseases with clinical implications. \citep{kiani2020impact} developed a deep learning-based system to aid pathologists in differentiating between two subtypes of primary liver cancer, hepatocellular carcinoma and cholangiocarcinoma, on H\&E stained whole slide images. \citep{Lu2020prognostic} demonstrated the usefulness of extracting image features from hepatocellular carcinoma histopathological images using the pre-trained CNN models to reliably differentiate between normal and cancer samples.

\subsection{Contributions}
\label{path_sec:path_contribs}

A deep learning-based framework for segmentation and analysis of whole slide images has been proposed. The framework comprises of segmentation network at its core along with novel algorithms that utilize the segmentation to do pathological analyses such as metastasis classification and viable tumour burden estimation. As discussed in section \ref{path_sec:1}, challenges in whole slide image analysis are mainly due to their large size, variability in staining, and the limited amount of annotated data. In this work, the following contributions were proposed for addressing the aforementioned problems:

\begin{itemize}
	\item Ensemble segmentation model: The ensemble comprised of multiple FCN architectures, each independently trained on different subsets of the training data. During inference, the ensemble generated the tumour probability map by averaging the posterior probability maps of all the FCNs. The ensemble approach showed superior segmentation performance when compared to its individual constituting FCNs.   

	\item Training pipeline: The proposed approach divided the whole slide images into smaller sized image patches for the purpose of training FCN models. For the preparation of the training set, efficient methodologies for sampling patches from the whole slide images were introduced. The problem of class-imbalance due to the limited number of representative patches from tumour regions in the whole slide images was addressed by employing overlapping and oversampling techniques during patch extraction (random patch coordinate perturbation technique) alongside with various data augmentation schemes.

	\item Inference pipeline: For efficiently generating model inference on the entire whole slide image, a concept of generating patch coordinate sampling grid from the post-processed tissue mask was introduced. The sampling grid aided in the reduction of computational time by discarding non-tissue patches during the construction of the tumour probability heatmap. The patch-based segmentation of whole slide images introduced edge artefacts due to loss of neighbouring context information at patch borders, and this issue was addressed by proposing techniques to average prediction probabilities at overlapping regions and making use of large patch size during inference. Apart from this, we also compute inference on multiple models parallelly for ensemble calculation over patches rather than an entire image.

	\item Lymph node metastases classification from whole slide images: A Random Forest-based ensemble classification algorithm was trained with hand-crafted features derived from the predicted tumour-probability maps. The class-imbalance in the training dataset was addressed by employing strategies such as over-sampling (by synthetically generating under-presented class data points) and under-sampling (balance all classes by removing some noisy data points).


    
    
    \item Uncertainty estimation: An efficient patch-based uncertainty estimation framework was developed to estimate both data specific and model (parameter) specific uncertainties.

    \item Open-source Packaging:  The proposed framework was packaged into an open-source GUI application for the benefit of researchers.
   
\end{itemize}
The performance of the segmentation pipeline was benchmarked by validating it on whole slide images of three different cancer sites namely- breast lymph nodes, liver and colon by participating in CAMELYON \citep{litjens20181399}, DigestPath \citep{li2019signet}, and PAIP citep{paip} challenges respectively. The framework is packaged into an open-source GUI application for the benefit of researchers \footnote{https://github.com/haranrk/DigiPathAI}.

\section{Materials and methods}
\subsection{Overview}
\begin{figure*}
    \includegraphics[width=.7\textwidth]{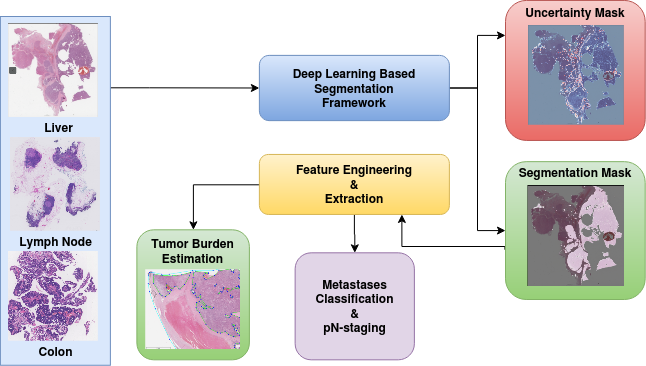}
    \centering
    \caption{Deep learning based framework for segmentation and analysis of whole slide images.}
    \label{path_fig:overview}
\end{figure*}
The Figure \ref{path_fig:overview} provides an overview of the proposed deep learning based segmentation and downstream analyses framework for whole slide images corresponding to multiple different cancer sites.

\subsection{Datasets used for this study}
\label{path_sec:path_data}
The proposed framework was validated on multiple open-source datasets which included CAMELYON \citep{litjens20181399}, PAIP \citep{paip} and DigestPath \citep{li2019signet}. Table \ref{path_tab:datasets} provides an overview of the datasets used in this study.

\begin{table}
\centering
\caption{Summary of histopathological datasets used in this work. The test images were hidden by the competition organizers and used only for leaderboard evaluation.}
\label{path_tab:datasets}
\setlength{\tabcolsep}{3pt}
\begin{tabular}{@{}lllll@{}}
  
Dataset     & Train set     & Test set  & $<Image Size>$ &  Microns  \\
            &               &           &                &  /pixel    \\    
CAMELYON16                                    & 270                        & 129                       & 100,000x100,000                     & 0.25                                                                                       \\
CAMELYON17                                    & 500                        & 500                       & 100,000x100,000                     & 0.25                                                                                      \\
DigestPath                                    & 660                        & 212                       & 5,000x5,000                         & 0.25                                                                                       \\
PAIP                                          & 50                         & 40                        & 50,000x50,000                       & 0.5                                                                                        \\  
\end{tabular}
\end{table}

%

%

\subsubsection{CAMELYON16}

\begin{table}
\centering
\caption{Tumour size criteria for assigning metastasis type.}
\label{path_tab:meta_size}
\begin{tabular}{@{}ll@{}}
  
\textbf{Category}             & \textbf{Size}                                                                                                                                     \\    
Isolated tumour cells & \begin{tabular}[c]{@{}l@{}}Single tumour cells or a cluster of\\ tumour cells not larger than 0.2\\ mm or less than 200 cells\end{tabular} \\
Micro-metastasis     & \begin{tabular}[c]{@{}l@{}}Larger than 0.2 mm and/or con-\\ taining more than 200 cells, but\\ not larger than 2 mm\end{tabular}         \\
Macro-metastasis     & Larger than 2 mm                                                                                                                         \\  
\end{tabular}
\end{table}

The CAMELYON16 \citep{bejnordi2017diagnostic} dataset comprised of 399 whole slide images taken from two medical centres in the Netherlands, out of which 159 whole slide images were metastases, and the remaining 240 were negative. Pathologists exhaustively annotated all the whole slide images with metastases at the pixel level. In the CAMELYON16 challenge, the 399 whole slide images were split into training and testing sets, comprising of 160 negative and 110 metastases whole slide images for training, 80 negative and 49 metastases whole slide images for testing.

\subsubsection{CAMELYON17}
The CAMELYON17\citep{bandi2018detection} dataset consisted of 1000 whole slide images taken from five medical centres in the Netherlands. In the CAMELYON17 challenge, 500 whole slide images were allocated for training, and the remaining 500 whole slide images were allocated for testing. The training dataset of CAMELYON17 included 318 negative whole slide images and 182 whole slide images with metastases. In CAMELYON17 dataset, slide-level labels of metastases type were provided for all the whole slide images, and exhaustive pixel-level annotations were provided for 50 whole slide images. The slide-level labels were negative, Isolated Tumor cells (ITC), micro-metastases and macro-metastases. Table \ref{path_tab:meta_size} provides the size criteria for metastases type. 
The pN-stage labels were provided for all the 100 patients in the training set and were based on the simplified rules provided in Table \ref{path_tab:pn-stage}. The Table \ref{path_tab:cm17-data} provides the metastases type distribution in CAMELYON17 training dataset.

\begin{table}
\centering
\caption{Metastases type distribution in CAMELYON17 training set.}
\label{path_tab:cm17-data}
\begin{tabular}{llll}
\hline
\multicolumn{4}{c}{Metastases (Training Set)} \\ \hline
Negative     & ITC     & Micro     & Macro    \\
318          & 35      & 64        & 88       \\ \hline
\end{tabular}%
\end{table}

\begin{table}
\centering
\caption{Pathologic lymph node classification (pN-stage) in CAMELYON17 Challenge.}
\label{path_tab:pn-stage}
\begin{tabular}{@{}ll@{}}
  
\textbf{pN-Stage} & \textbf{Slide Labels}                                                   \\    
pN0      & No micro-metastases or macro-metastases or \\
         & ITCs found.                            \\
pN0(i+)  & Only ITCs found.                                                                  \\
pN1mi    & Micro-metastases found, but no macro-metastases \\
         & found.                            \\
pN1      & Metastases found in 1-3 lymph nodes, of which at \\ 
         & least one is a macro-metastasis. \\
pN2      & Metastases found in 4-9 lymph nodes, of which at \\
         & least one is a macro-metastasis. \\  
\end{tabular}
\end{table}

\begin{table}
\centering
\caption{List of features extracted for the purpose of predicting lymph node metastases type. Features were extracted after thresholding tumour probability heatmaps. For feature numbers 5, 6, 7, 8 and 9 the following statistics were computed- maximum, mean, variance, skewness, and kurtosis.}
\label{path_tab:metastases-features}
\setlength{\tabcolsep}{3pt}
\begin{tabular}{@{}lll@{}}
  
\textbf{No.} & \textbf{Feature description}                                        & \textbf{Threshold (p)} \\    
1            & Largest tumour region’s major & \\ 
             & axis length                            & p=0.9 \& p=0.5         \\
2            & Largest tumour region's area                                         & p=0.5                  \\
3            & Ratio of tumour region to  & \\
            & tissue region                              & p=0.9                  \\
4            & Count of non-zero pixels                                            & p=0.9                  \\
5            & Tumour regions area         & p=0.9                  \\
6            & Tumour regions perimeter    & p=0.9                  \\
7            & Tumour regions eccentricity & p=0.9                  \\
8            & Tumour regions extent       & p=0.9                  \\
9            & Tumour regions solidity     & p=0.9                  \\
10           & Mean of all region's mean & \\
             & confidence probability                 & p=0.9                  \\
11           & Number of connected regions                                         & p=0.9                  \\  
\end{tabular}%
\end{table}

\subsubsection{PAIP}
\begin{figure}[!t]
    \centerline{\includegraphics[width=\columnwidth]{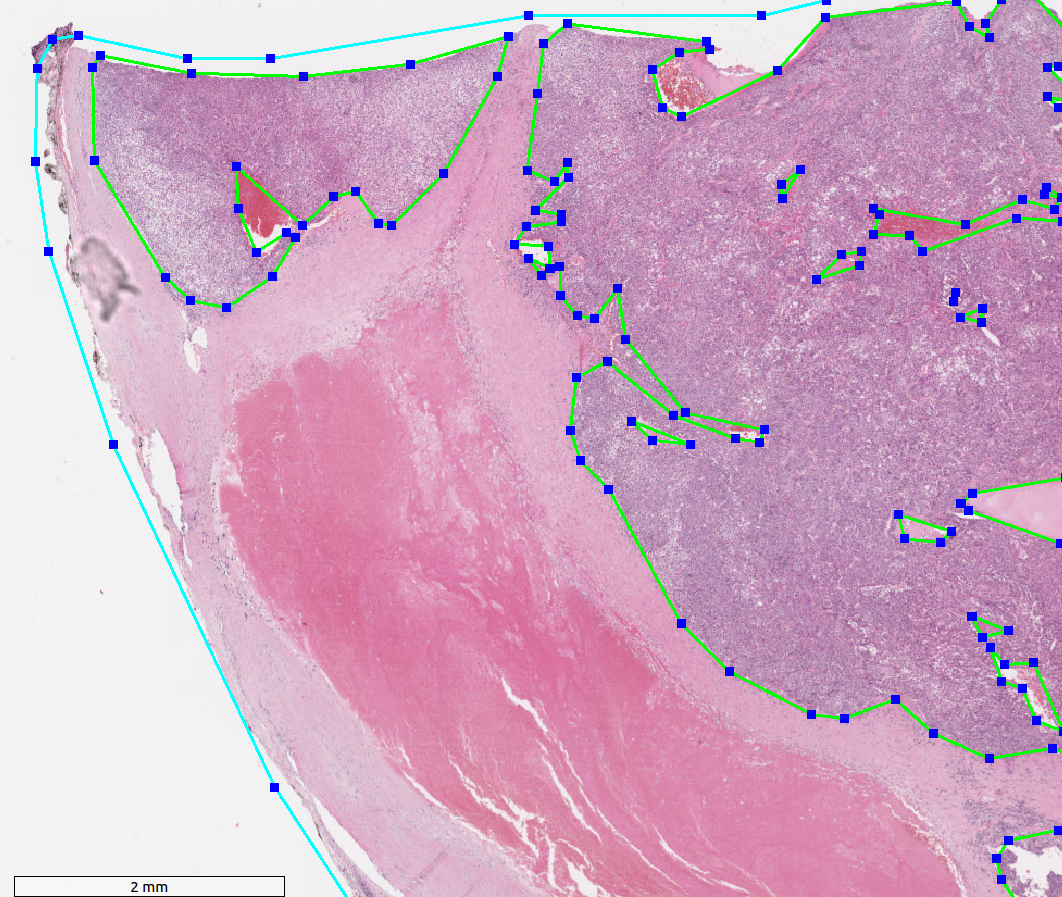}}
    \caption{Whole slide image of a liver tumour from the PAIP dataset. The green contour represents the viable tumour and the blue contour represents whole tumour. Annotations were made by expert pathologists.}
    \label{path_fig:viable-whole-tumour}
\end{figure}

The PAIP 2019 \citep{paip} dataset contains a total of 100 whole slide images scanned from liver tissue samples. Each image has an average dimension of 50,000x50,000 pixels. All the images were H\&E stained, scanned at 20x magnification and prepared from a single centre, Seoul National University Hospital. The dataset included pixel-level annotation of the viable tumour and whole tumour regions. It also provided the viable tumour burden metric for each image.

Tumour burden is defined as the ratio of the viable tumour region to the whole tumour region. The viable tumour region is defined as the cancerous region. The whole tumour area is defined as the outermost boundary enclosing all the dispersed viable tumour cell nests, tumour necrosis, and tumour capsule (Fig \ref{path_fig:viable-whole-tumour}). Each tissue sample contains only one whole tumour region. This metric has applications in assessing the response rates of patients to cancer treatment.

Out of the 100 images, 50 images were the publicly available training set, ten images were reserved for validation set that was made publicly available after the challenge was completed, and the rest 40 images were the test set whose ground truth were not publicly available.

\subsubsection{DigestPath}
DigestPath dataset consists of tissue sections collected during the examination of colonoscopy pathology to identify early-stage colon tumour cells. There are ten or more tissues sections in a single whole slide image for colonoscopy pathology review. The challenge organisers selected one or two tissue sections in a whole slide image and provided images of these tissue sections along with their corresponding lesion annotations by pathologists in jpg format. On average, each tissue image was of size 5000x5000 pixels. The training dataset of DigestPath consists of 660 tissue images taken from 324 patients, in which 250 tissue images from 93 patients had lesions, and the remaining 410 tissue images from 231 patients had no lesions. The data was collected from multiple medical centres, especially from several small centres in developing countries. All the tissues sections were H\&E stained and scanned at 20x magnification. The testing dataset consisted of 212 tissue images from 152 patients. The challenge organisers released only the training set and the testing set were kept confidential. 

\subsection{Data pre-processing}
\label{path_sec:path_datapreprocess}

\subsubsection{Tissue mask generation}
\label{path_sec:tm}
In this step, the entire tissue region was segmented from the background glass region of the whole slide image. This step aided in preventing unnecessary computations on non-tissue regions of the slide. An approximate tissue region boundary suffices, therefore the processing was done on a low resolution version of the whole slide image to further reduce computational costs. The RGB colour space of the low-resolution image was transformed to HSV (Hue-Saturation-Value) colour space and Otsu's adaptive thresholding \citep{otsu1979threshold} was applied to the saturation component. Post thresholding, binary morphological operations were performed to facilitate proper extraction of patches at the small tissue regions and tissue borders.
\subsubsection{Tissue mask generation specific to CAMELYON dataset}
\begin{figure}
    \includegraphics[width=1.\columnwidth]{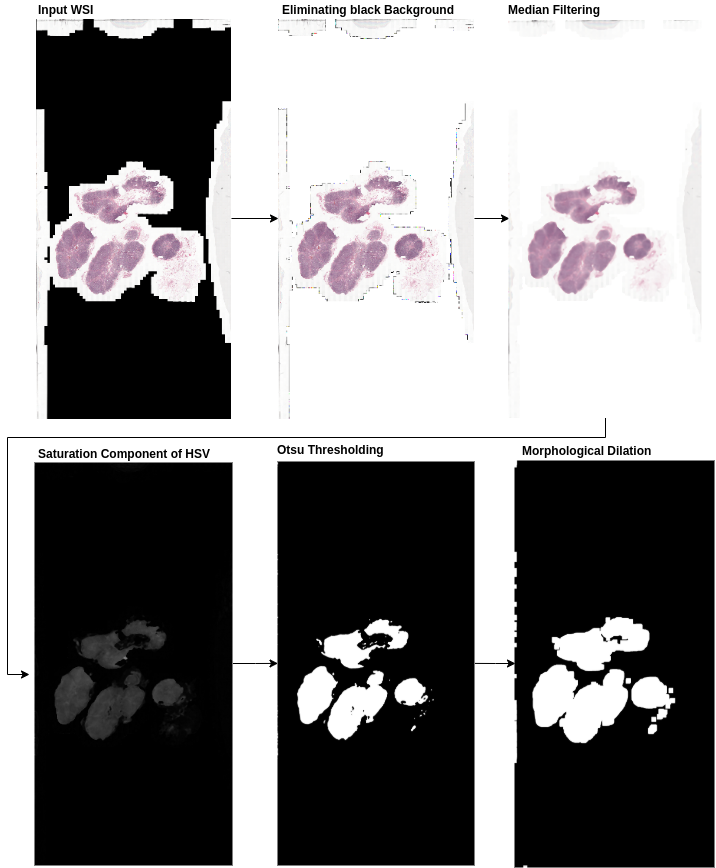}
    \caption{An illustration of the intermediate stages in the process of tissue mask generation from a whole slide image in CAMELYON17 dataset.}
    \label{path_fig:cm17_tmask}
\end{figure}

In some of the CAMELYON17 cases, the Otsu's thresholding failed because of the black regions in the whole slide image. Hence, before the application of image thresholding operation, the pre-processing involved replacing black pixel regions in the whole slide image back-ground with white pixels and median blurring with a kernel of size 7x7 on the entire image. Median blur aided in the smoothing of the tissue regions and removal of noise at the tissue bordering the glass-slide region while preserving the edges of the tissue. Figure \ref{path_fig:cm17_tmask} illustrates the pipeline for tissue mask generation with an example.

\subsubsection{Patch coordinate extraction}
Using the tissue mask, patches of the image were randomly extracted to make the training dataset. An equal number of tumorous and non-tumorous patches were extracted. A patch was considered tumorous if at least one pixel inside the patch was classified a tumor. The dimensions of the extracted patches were not fixed. Rather, they were a hyperparameter we experimented with. In this step, only the centers of the potential training patches were extracted and stored.




\subsubsection{Data augmentation}
To increase the number of data points and to better generalize the models across various staining and acquisition protocols, data augmentation schemes were proposed. Augmentations like  “horizontal or vertical flip,” “90-degree rotations”, and “Gaussian blurring” along with colour augmentation were performed. Colour augmentation included random changes to brightness, contrast, hue, and saturation with a maximum delta of 64.0/255, 0.75, 0.25, 0.04, respectively.

Additionally, in order to introduce some diversity of patches extracted from the whole slide images during each training epoch, random coordinate perturbation was introduced. The random coordinate perturbation involved offsetting the centre of the patch coordinates within a specified radius prior to the extraction from the whole slide image. The radius was fixed to be 128 pixels, and patches of size 256x256 pixels were extracted from the highest resolution image. Since in the patch extraction phase, only the centers were extracted and stored, this augmentation could be applied on the fly during training. Post augmentation, the images were normalized.



\subsection{Network architecture}
\label{path_sec:path_networks}

For the task of segmentation of tumour regions from patches of the whole slide images fully convolutional neural network (FCN) \citep{long2015fully}  based architectures were used. A typical FCN based segmentation network comprises of an encoder network, a decoder network and a pixel-wise classification layer. An encoder network comprises of a series of operations (like convolution and pooling) that transforms the input (image) to a set of low resolution feature maps. The decoder network comprises of up-sampling or transposed convolution followed by series of convolution operations that transform the low resolution encoder feature maps to the original input resolution feature maps for pixel-wise classification. 

The ensemble consisted of three encoder-decoder based FCN architectures. During inference, the predicted tumour posterior probability map from all the three models were averaged to generate the ensemble model's final prediction. The ensemble comprised of the following FCN architectures:
\begin{itemize}
    \item U-Net \citep{ronneberger2015u} with DenseNet-121 \citep{huang2017densely} as the backbone (encoder) pre-trained on ImageNet \citep{deng2009imagenet}. The decoder comprised of the bi-linear up-sampling module followed by convolutional layers. Features learnt in the down-sampling path of the encoder were concatenated with the features learnt in the up-sampling path using skip connection.

    \item U-Net \citep{ronneberger2015u} with Inception-ResNet-V2 \citep{szegedy2017inception} as the backbone (encoder) pre-trained on ImageNet \citep{deng2009imagenet}. The Inception-ResNet-V2 \citep{szegedy2017inception} (also known as Inception-v4) integrates the features of the Inception architecture \citep{szegedy2015going} and the ResNet architecture \citep{he2016deep}.

    \item DeeplabV3Plus \citep{chen2018encoder} with Xception \citep{chollet2017xception} network as the backbone and pre-trained on PASCAL VOC \citep{everingham2010pascal}. DeepLabV3 \citep{chen2017rethinking} was built to obtain multi-scale context. This was done by using atrous convolutions with different rates. DeeplabV3Plus extends this by having low-level features transported from the encoder to decoder.
\end{itemize}

\subsection{Loss function}
Tumour regions were represented by a minuscule proportion of pixels in whole slide images, thereby leading to class imbalance. This issue was circumvented by training the network to minimize a hybrid loss function. The hybrid loss function comprised of cross-entropy loss and a loss function based on the Dice overlap coefficient. The Dice coefficient is an overlap metric used for assessing the quality of segmentation maps. 
The dice loss is a differentiable function that approximates Dice-coefficient and is defined using the predicted posterior probability map and ground truth binary image as defined in \ref{diceloss}. The cross-entropy loss is defined in \ref{crossentropyloss}. In the equations, $p_i$ and  $g_i$  represent pairs of corresponding pixel values of predicted posterior probability and ground truth. $N$ represents the total number of pixels. $DL$ refers to dice loss and $CL$ refers to cross-entropy loss. $DL_{FG}$ and $DL_{BG}$ represent the foreground pixels that correspond to the tumour regions and the background pixels that corresponded to non-tumour regions, respectively.

\begin{equation}
    DL = 1 - \frac{2\sum_i ^N p_i g_i}{ \sum_i^N p_i^2+ \sum_i^N g_i^2}
    \label{diceloss}
\end{equation}
\begin{equation}
    CL = \sum_i ^N \left(g_ilog(p_i) + (1 - g_i)log(1-p_i)\right)
    \label{crossentropyloss}
\end{equation}
\begin{equation}
 Loss = \alpha*CL + \beta*DL_{BG} + \gamma*DL_{FG} 
 \label{combinedloss}
\end{equation}

The proposed loss was defined as a linear combination of the two-loss components as defined in \ref{combinedloss}. The neural networks were trained by minimizing the proposed loss function using ADAM optimizer (\citep{kingma2014adam}). The $\alpha, \beta, \gamma$  were empirically assigned to the individual loss components.  The following configurations were set: $ \alpha = 0.5, \beta = 0.25$ and $\gamma = 0.25$.

\subsection{Uncertainty analysis}
\label{path_sec:path_uncertainty}
Uncertainty estimation is essential in assessing unclear diagnostic cases predicted by deep learning models. It helps pathologists to concentrate more on the uncertain regions for their analysis.  \citep{begoli2019need} argues the need for uncertainty analysis in machine-assisted medical decision-making system. There exist two main sources of uncertainty, namely (i) Aleatoric uncertainty and (ii) Epistemic uncertainty. Aleatoric uncertainty is uncertainty due to the data generation process itself. In contrast, the uncertainty induced due to the model parameters, which is the result of not estimating ideal model architectures or weights to fit the given data, is known as epistemic uncertainty \citep{kendall2017uncertainties}. Epistemic uncertainty can be approximated by using test time Bayesian dropouts as discussed in \citep{leibig2017leveraging}, which estimates uncertainty by Montecarlo simulations with Bayesian dropout. 

In the proposed pipeline, aleatoric uncertainty for each model was estimated using test time augmentations, as introduced in \citep{gal2016dropout} \ref{eq:aleatoric}. 
\begin{equation}
 var_{al}(x, \Phi_i) \approx \mathbf{E}_{t \sim TTA}[(\Phi_i(x|w,t) -  \mathbf{E}_{t \sim TTA}[\Phi_i(x|w,t)])^2]
\label{eq:aleatoric}
\end{equation}
where $\Phi_i(x|w)$ is the output of the neural network with weights $w$ for input $x$ and $TTA$ denotes the set of possible test time data augmentations allowed. The proposed methodology for aleatoric uncertainty included the following augmentations- $TTA \in \{rotation, vertical flip, horizontal flip\}$.

For epistemic uncertainty, the diversity of model architectures were used to calculate uncertainty \ref{eq:epistemic}. 
\begin{equation}
 var_{ep}(p(y|x,w)) \approx \mathbf{E}_{\phi \sim \{\Phi_i\}}[(\phi(x|w) - \mathbf{E}_{\phi \sim \{\Phi_i\}}[\phi(x | w)])^2]
\label{eq:epistemic}
\end{equation}
where the likelihood distribution $p(y|x,w)$ is a probabilistic model which generates outputs ($y$) for given inputs ($x$) for some parameter setting ($w$) and $\Phi_i$ indicates the trained model.

\subsection{Training pipeline}
\label{path_sec:train_pipeline}
\begin{figure*}
    \centering
    \includegraphics[width=\textwidth]{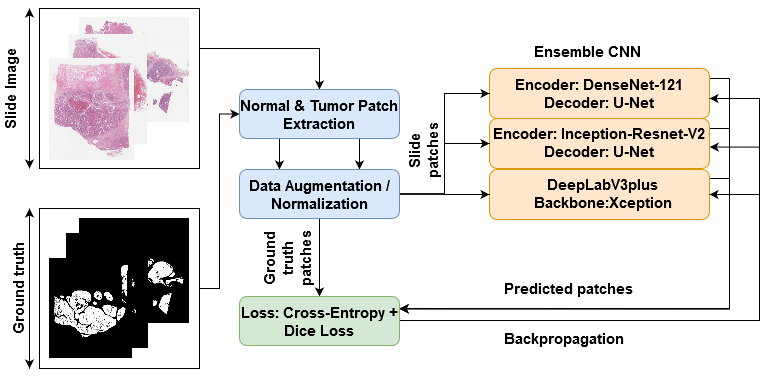}
    \caption{Overview of the tumour segmentation training pipeline.}
    \label{path_fig:training_pipeline}
\end{figure*}

Figure \ref{path_fig:training_pipeline} illustrates the training strategy utilized for training each of the models in the ensemble. The batches for training were generated with an equal number of tumour and non-tumour patches. This was done to prevent class imbalance or manifold shift issues and enforce proper training. All three models were trained independently, with different cross-validation folds of the data. The FCN architectures in the ensemble whose encoders were based on DenseNet-121 and Inception-ResNet-V2 made use of transfer learning by using ImageNet \citep{deng2009imagenet} pre-trained weights for their respective encoders. In the case of DeeplabV3Plus, the model weights were pre-trained on PascalVOC \citep{everingham2010pascal}. For the network architectures with encoders based on DenseNet-121 and Inception-ResNet-V2, the encoder weights of the models were frozen for the first two epochs, and only the decoder weights were made trainable. For the remaining epochs, both the encoder and decoder parts were trained. The learning rate was decayed every few epochs in a deterministic manner to allow for the model to gradually converge.

\subsection{Inference pipeline}
\label{path_sec:inference_pipeline}

\begin{figure}
    \includegraphics[width=\columnwidth]{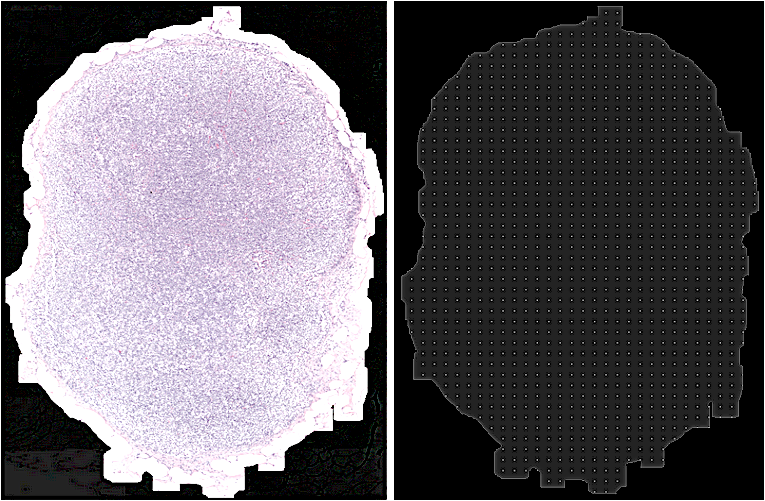}
    \caption{(Left to Right) An illustration of the tissue mask overlayed on a small region of the whole slide image at low resolution (level-4), here the white region corresponds to the tissue mask; An illustration of the generated uniform patch coordinate sampling grid, here the points on the image act as centres from which high-resolution image patches were extracted from the whole slide image.}
    \label{path_fig:patch_grid}
\end{figure}

The pre-processing step in the inference pipeline included segmentation of tissue region from the whole slide image (refer \ref{path_sec:tm}).  In order to facilitate extraction of patches from the whole slide image within the tissue mask region, a uniform patch-coordinate sampling grid was generated at a lower resolution, as shown in Figure \ref{path_fig:patch_grid}. Each point in the patch sampling grid was re-scaled by a factor to map to the coordinate space corresponding to the whole slide image at its highest resolution. From these scaled coordinate points as the centre, fixed-size high-resolution image patches were extracted from the whole slide image for feeding the trained segmentation model's input. The sampling stride was defined as the spacing between consecutive points in the patch sampling grid. The patch size and the sampling stride controlled the overlap between consecutive extracted patches from the whole slide image. The main drawback of patch-based segmentation method for whole slide image was that the smaller patch sizes could not capture wider context of the neighbourhood region. Moreover, stitching of the segmented patches introduced boundary artefacts (blockish appearance) in the tumour probability heatmaps. The generated heatmaps were smooth when the inference was done on overlapping patches with larger patch-size and averaging the prediction probabilities at the overlapping regions. The experimental observation suggested that 50\% overlap between consecutive neighbouring patches was the optimal choice as it ensured that a particular pixel in a whole slide image was seen at most 4-times during the heatmap generation. However, this approach increased the inference time by a factor of 4. Also, during inference, increasing the patch size by a factor of 4 (1024x1024) when compared to the patch size used during training (256x256) improved the quality of generated heatmaps. 

\subsection{pN-staging pipeline for CAMELYON17 dataset}
\label{path_sec:path_pnstaging}

\begin{figure}[!h]
    \centering
    \includegraphics[width=1.\columnwidth]{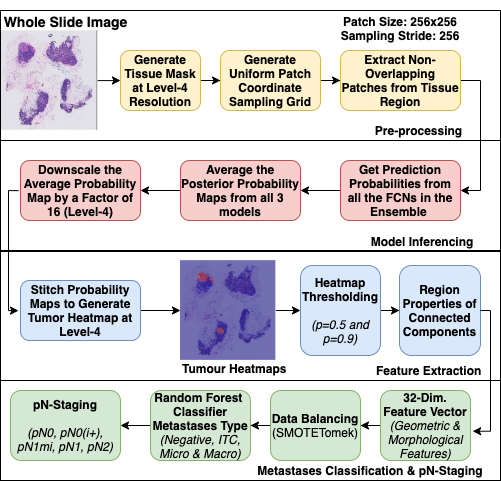}
    \caption{Overview of the steps involved in the pN-staging pipeline developed for CAMELYON17 dataset.}
    \label{path_fig:pn-pipeline}
\end{figure}

Figure \ref{path_fig:pn-pipeline} illustrates the complete pipeline developed for pN-staging of CAMELYON17 dataset. The pipeline comprises four blocks as described below:
\begin{itemize}
    \item Pre-processing: The tissue regions in the whole slide images were detected for patch extraction.
    \item Heatmap generation: The extracted patches from the whole slide images were passed through the inference pipeline to generate the down-scaled version of the tumour probability heatmaps.
    \item Feature extraction: The heatmaps were binarized by thresholding at 0.5 and 0.9 probabilities, and at each of these thresholds, the connected components were extracted, and region properties were measured using scikit-image \citep{scikit-image} library. Thirty-two geometric and morphological features from the probable metastases regions were computed.

    \item Data balancing: In order to handle the class imbalance problem, one of the techniques proposed in the literature is oversampling by synthetically generating minority class samples using SMOTE algorithm \citep{chawla2002smote}. However, this method can introduce noisy samples when the interpolated new samples lie between marginal outliers and inliers. This problem is usually addressed by removing noisy samples by using under-sampling techniques like Tomek's link \citep{tomek1976two} or nearest-neighbours. SMOTETomek \citep{batista2004study} algorithm was employed for balancing the training data. SMOTETomek algorithm is a combination of SMOTE and Tomek's link performed consecutively.    
    
    \item Classification: The pN-stage was assigned to the patient based on all the available lymph-node whole slide images, taking into account their individual metastases type (Table \ref{path_tab:meta_size}). For predicting the metastases type, an ensemble of Random Forest classifiers \citep{liaw2002classification} was trained using the extracted features.
\end{itemize}

\subsection{Tumour burden estimation for PAIP dataset}
\label{path_sec:path_tumorburden}

The tumour burden computation requires the segmentation of viable tumour and whole tumour regions in the whole slide image of the liver cancer tissue. The viable tumour region was segmented using the proposed deep learning-based segmentation network. However, it was observed that training the same segmentation network for whole tumour region gave sub-optimal results. Hence, a heuristic method was adopted to approximate the whole tumour region from viable tumour region.

The tumour burden estimation algorithm consisted of the following steps:
\begin{itemize}
    \item Segment the viable tumour region from image
    \item Apply morphological operation to remove false positives and fill holes
    \item Find the smallest convex hull containing the entire viable tumour region
    \item Estimate the tissue mask, as discussed in \ref{path_sec:tm}.  
    \item The whole tumour region is approximated to be the intersection of the convex hull and tissue mask region.
    \item The tumour burden is calculated by taking the ratio between the area of the viable and whole tumour regions
\end{itemize}

\section{Experimental analysis}
\label{path_sec:expi_analysis}
In this section, the effectiveness of the proposed methodologies for segmentation and classification models are experimentally analyzed. The neural networks were implemented using TensorFlow-Keras (\citep{abadi2016tensorflow}) deep learning framework. The experiments were run on multiple desktop computers with NVIDIA Titan-V GPU with 12 GB RAM, Intel Core i7-4930K 6-core CPUs @ 3.40GHz, and 48GB RAM.

\subsection{Lesion detection analysis on CAMELYON16 dataset}
In this section, some of the techniques specific to CAMELYON dataset pre-processing are detailed, and discussion on the performance of various FCN architectures and ensemble configurations for lesion detection on the CAMELYON16 test dataset (n=139) are provided.

\subsubsection{FROC evaluation score}
\label{pth_sec:FROC}
One of the metrics used in CAMELYON16 challenge for lesion-based evaluation is free-response receiver operating characteristic (FROC) curve. The FROC curve is defined as the plot of sensitivity versus the average number of false-positives per image. The CAMELYON16 challenge testing dataset was used for evaluating the performance of the proposed algorithms for lesion detection/localization. The detection/localization performance was summarized using Free Response Operating Characteristic (FROC) curves. This was similar to ROC analysis, except that the false positive rate on the x-axis is replaced by the average number of false positives per whole slide image. The following guidelines were followed for lesion detection in CAMELYON16 challenge.
\begin{itemize}
    \item If the position of the detected region was inside the annotated ground truth lesion it was considered a true positive 
    
    \item If a single ground-truth region had several findings, they were counted as a single true positive finding, and none of them were counted as false positives 
    
    \item All detections which were not within a reasonable distance of the annotations of ground truth were counted as false positives
    
    \item The final FROC score was defined as the average sensitivity at six predefined false positives: 1/4, 1/2, 1, 2, 4 and 8 FPs per whole slide image 
\end{itemize}


\subsubsection{Dataset preparation specific to CAMELYON dataset}
For training the ensemble segmentation model for lesion segmentation, training sets of both CAMELYON16 and CAMELYON17 dataset which had pixel-level annotations for the whole slide images were used. As noted by the challenge organizers, some of the whole slide images were not exhaustively annotated in the CAMELYON16 training set; such slides were excluded in training set preparation. So, in total, 628 whole slide images for training were utilized (250 whole slide images from CAMELYON16 and 378 from CAMELYON17). A three-fold stratified cross-validation split of the training set was done to maximize the utilization of the limited number of whole slide images. The stratification ensured that the ratio of negative to metastases was maintained in all the three folds. From 628 whole slide images, 5,71,029 coordinates whose patches corresponded to regions from the tumour and non-tumour tissue regions were randomly sampled. A patch extracted from a whole slide image was labelled as a tumour patch if it had non-zero pixels labelled as tumour pixels in the pathologist’s manual annotation. Further, these extracted patch coordinates were distributed into their respective cross-validation folds. Table \ref{path_tab:fold_dist}  shows the distribution of the split in each of the folds.

\begin{table}
\centering
\caption{Count of the tumour and non-tumour patches in each of the three cross-validation folds.
}
\label{path_tab:fold_dist}
\begin{tabular}{@{}llll@{}}
  
\textbf{Patch label} & \multicolumn{3}{l}{\textbf{No. of patch coordinates}} \\    
                     & \textbf{Fold-0}          & \textbf{Fold-1}          & \textbf{Fold-2}         \\
Training Non-Tumour   & 1,87,034           & 1,96,094           & 1,90,424          \\
Training Tumour       & 1,84,467           & 1,94,709           & 1,87,440          \\
Validation Non-Tumour & 99,742            & 90,682            & 96,352           \\
Validation Tumour     & 99,786            & 89,544            & 96,813           \\  
\end{tabular}%
\end{table}

\subsubsection{Training and inference configuration of ensemble FCN models}
\label{path_sec:ens_config}
\begin{figure}
    \includegraphics[width=1.\columnwidth]{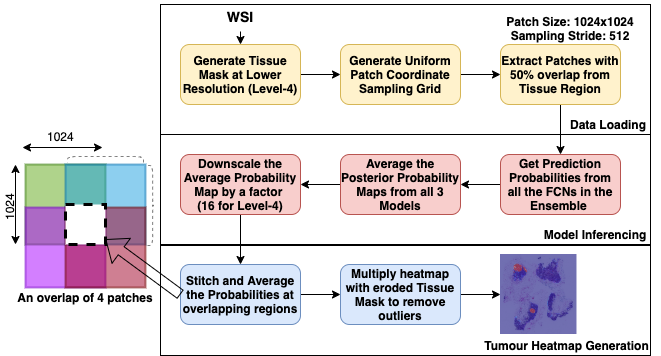}
    \caption{The figure illustrates the overlap-stitch inference pipeline used in Ensemble-B configuration.}
    \label{path_fig:EnsB-config}
\end{figure}

The following two ensemble configurations were proposed:  
\begin{itemize}
    \item Ensemble-A: Comprised of the three different FCN architectures, as described in section \ref{path_sec:train_pipeline}. The inference pipeline made use of patch size of 256 and extracted non-overlapping patches.
    \item Ensemble-B: Comprised of three replicated versions of a single FCN architecture. The inference pipeline made use of a patch size of 1024 with a 50\% overlap between neighbouring patches, as illustrated in Figure \ref{path_fig:EnsB-config}.    
\end{itemize}

In both the ensemble configurations, each model in the ensemble was trained with one of the 3-fold cross-validation splits. All the models made use of pre-trained weights with the fine-tuning procedure, as described in section \ref{path_sec:train_pipeline}. The models were trained for ten epochs with a batch size of 32.

\subsubsection{Lesion detection performance of Ensemble-A}

\label{path_sec:Ens-A-perform}
\begin{figure}
    \subfloat[]{\includegraphics[width=.5\columnwidth, keepaspectratio]{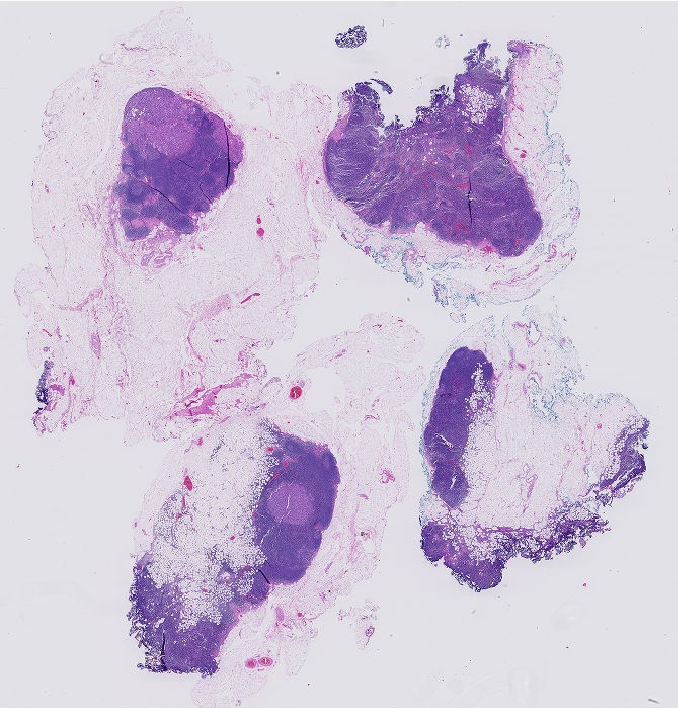}}
    \subfloat[]{\includegraphics[width=.5\columnwidth, keepaspectratio]{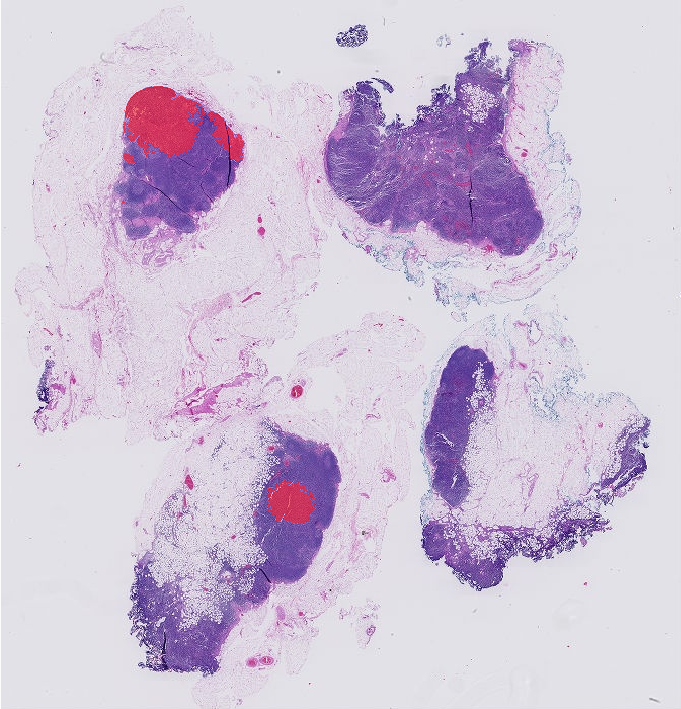}}
    \caption{(a) A whole slide image from CAMELYON16 test set, (b)  Tumour ground truth overlayed on the whole slide image.}
    \label{path_fig:Test_001_GT}
\end{figure}

\begin{figure}
    \includegraphics[width=1.\columnwidth]{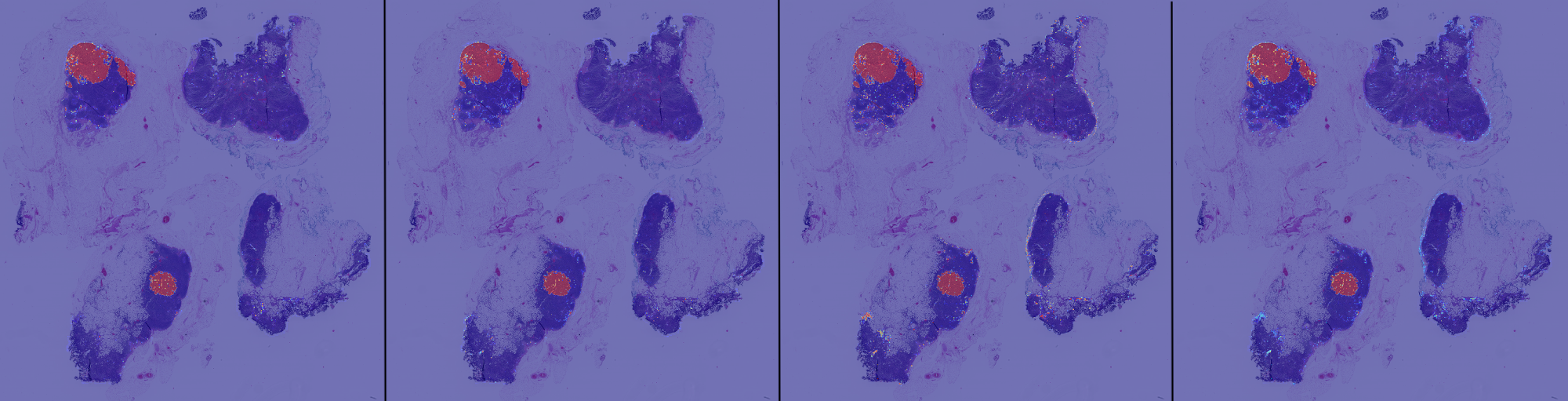}
    \caption{The figure shows the heatmaps overlayed on the whole slide image by FCN models in Ensemble-A configuration. (Left to Right): DenseNet-121 FCN, Inception-ResNet-V2 FCN, DeepLabV3plus, Ensemble, (Patch Size: 256x256, Sampling Stride: 256 pixels).}
    \label{path_fig:EnsA}
\end{figure}

Experimental results suggested that DenseNet-121 architecture had higher sensitivity and reduced false positives when compared to other FCNs in the ensemble configuration. It was also observed that Ensemble-A showed a significant difference in the FROC score compared to its constituents. The reason for this significant boost in the performance of Ensemble-A could be attributed to the effect of averaging the heatmaps from multiple FCN models, thereby lowering the probabilities of uncertain or less confident regions and hence eliminating the false positives. Figure \ref{path_fig:EnsA} illustrates the heatmaps generated by Ensemble-A and its constituent FCN models on a CAMELYON16 test case (refer Figure \ref{path_fig:Test_001_GT}). 

\begin{table}[!h]
\centering
\caption{FROC scores achieved on CAMELYON16 test set (n=139) by FCN models in Ensemble-A configuration. Note the abbreviations: IRF - Inception-ResNet-V2 FCN, DF- DenseNet-121 FCN, DL- DeepLabV3Plus, FP- false positives.}
\label{path_tab:EnsA-FROC}
\setlength{\tabcolsep}{3pt}
\begin{tabular}{@{}lllll@{}}
  
\textbf{Avg. FPs} & \\
    \textbf{/Slide} &  \multicolumn{4}{c}{\textbf{Sensitivity}}\\    
\textbf{}            & \multicolumn{1}{c}{\textbf{IRF(Fold-0)}} & \multicolumn{1}{c}{\textbf{DF(Fold-1)}} & \multicolumn{1}{c}{\textbf{DL(Fold-2)}} & \multicolumn{1}{c}{\textbf{Ensemble-A}} \\
0.25                      & 0.5                                       & 0.59                                     & 0.03                                     & 0.77                                    \\
0.5                       & 0.59                                      & 0.65                                     & 0.06                                     & 0.8                                     \\
1                         & 0.69                                      & 0.72                                     & 0.2                                      & 0.83                                    \\
2                         & 0.77                                      & 0.79                                     & 0.48                                     & 0.84                                    \\
4                         & 0.8                                       & 0.83                                     & 0.64                                     & 0.86                                    \\
8                         & 0.82                                      & 0.85                                     & 0.77                                     & 0.86                                    \\
\textbf{FROC} &&& \\
\textbf{Score}       & 0.69                                      & 0.74                                     & 0.36                                     &\textbf{0.83}                                    \\  
\end{tabular}%
\end{table}

\subsubsection{Lesion detection performance of Ensemble-B}
\label{path_sec:Ens-B-perform}

Since there was a marginal difference in FROC scores between various ensemble configurations (Table \ref{path_tab:froc-hyperparam}); hence in the interest of minimizing the computation time, Ensemble-A configuration was preferred with patch-size of 256x256 and sampling stride set to 256 (non-overlapping patch acquisition) for running inference on CAMELYON17 testing dataset (n=500). 

\begin{table}[!h]
\centering
\caption{FROC scores achieved on CAMELYON16 test set (n=139) by FCN models in Ensemble-B configuration (patch size- 1024 and sampling stride- 512). Note the abbreviations: DF-DenseNet-121 FCN, FP-false positives, F-Fold.}
\label{path_tab:EnsB-FROC}
\begin{tabular}{@{}lllll@{}}
  
\textbf{Avg. FPs} &\\
\textbf{/Slide} & \multicolumn{4}{c}{\textbf{Sensitivity}}                                                                                                                                \\    
\textbf{}                 & \multicolumn{1}{c}{\textbf{DF (F-0)}} & \multicolumn{1}{c}{\textbf{DF (F-1)}} & \multicolumn{1}{c}{\textbf{DF (F-2)}} & \multicolumn{1}{c}{\textbf{Ensemble-B}} \\
0.25                      & 0.56                                     & 0.56                                     & 0.61                                     & 0.77                                    \\
0.5                       & 0.65                                     & 0.63                                     & 0.69                                     & 0.84                                    \\
1                         & 0.71                                     & 0.70                                     & 0.76                                     & 0.85                                    \\
2                         & 0.77                                     & 0.75                                     & 0.81                                     & 0.88                                    \\
4                         & 0.82                                     & 0.80                                     & 0.84                                     & 0.88                                    \\
8                         & 0.86                                     & 0.86                                     & 0.88                                     & 0.89                                    \\
\textbf{FROC} &&&\\
\textbf{Score}       & 0.73                                     & 0.72                                     & 0.77                                     & \textbf{0.85}                           \\  
\end{tabular}%
\end{table}

\begin{table}
\centering
\caption{The table shows the FROC scores on CAMELYON16 test set (n=139) for various configurations of model, patch-size, and sampling-stride.}
\label{path_tab:froc-hyperparam}
\begin{tabular}{@{}llll@{}}

\textbf{Model} & \textbf{Patch size} & \textbf{Sampling stride} & \textbf{FROC score} \\    
Ensemble-A     & 256                 & 256                      & 0.83                \\
Ensemble-B     & 256                 & 256                      & 0.84                \\
Ensemble-B     & 1024                & 1024                     & 0.86                \\
Ensemble-B     & 1024                & 512                      & 0.85                \\  
\end{tabular}%
\end{table}

\begin{figure}
    \includegraphics[width=1.\columnwidth]{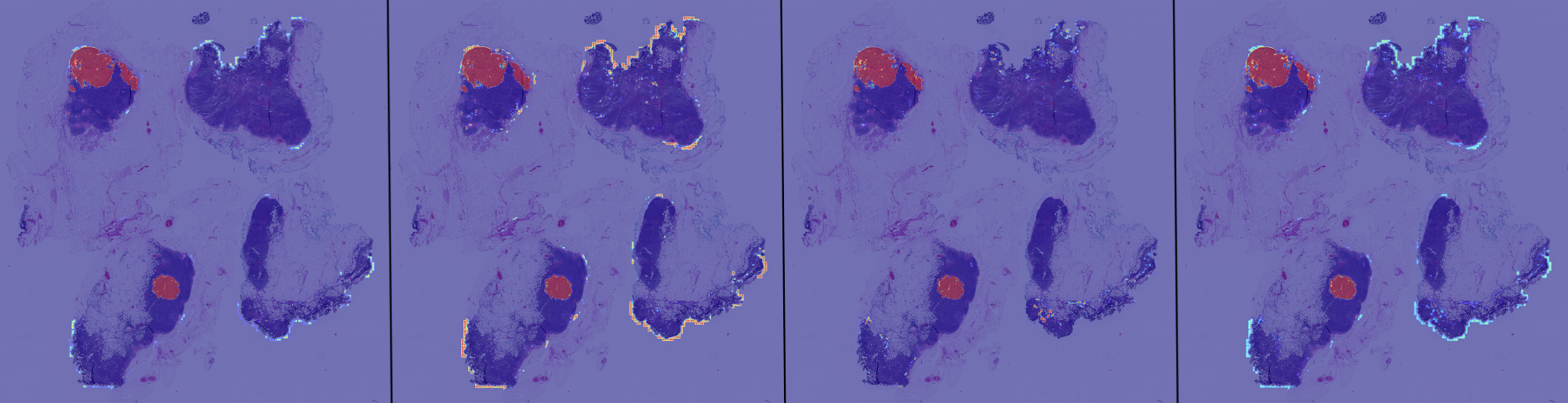}
    \caption{The figure shows the heatmap overlayed on the whole slide image by FCN models in Ensemble-B configuration. (Left to Right): 3 DenseNet-121 FCN models trained on cross-validation folds: 1, 0 and 2 respectively and Ensemble-B, (Patch Size: 1024x1024, Sampling Stride: 512 pixels). It can be seen from the heatmaps that the models struggled (relatively high posterior probability values at non-tumour regions) at extended regions from tissue boundaries.}
    \label{path_fig:EnsB}
\end{figure}

\subsection{Lymph-node metastases type classification analysis on CAMELYON17 dataset}
In this section, the experimental analysis of the classification model for lymph node metastases types is presented. 

\subsubsection{Cohen's kappa evaluation score}
Cohen's kappa \citep{fleiss1973equivalence} is a statistic that measures the inter-rater reliability for categorical variables. In CAMELYON17 challenge for evaluating pN-staging of the patients, the metric used was Cohen's kappa with five classes and quadratic weights. The kappa metric ranges from -1 to +1, where 1 represented perfect agreement with the raters, and 0 represented the amount of agreement that can be expected by random chance and, a negative value represented lower than chance agreement. 

\subsubsection{Dataset preparation}
The CAMELYON17 training dataset had 100 patients, and each patient had five whole slide images with their corresponding metastases labels (total 500 slide images).  The training dataset comprising of 100 patients were split into 43 patients as a train set and the remaining 57 patients as a validation set. The split ensured that the patients in the train set had at least one whole slide image with pixel-level annotation. The Table \ref{path_tab:cm17-datasplit} shows the distribution of whole slide images in terms of metastases type between train and validation sets; the proposed split strategy ensured that the distribution of metastases type between the two splits was similar.

\begin{table}
\centering
\caption{Metastases type distribution in CAMELYON17 train set and validation set.}
\label{path_tab:cm17-datasplit}
\begin{tabular}{@{}llllll@{}}
  
\textbf{Dataset} & \multicolumn{5}{c}{\textbf{No. of whole slide images}}\\
                 &  \multicolumn{5}{c}{\textbf{per each metastasis type}} \\    
                 & Negative       & ITC       & Micro      & Macro      & Total      \\
Train set        & 100            & 26        & 35         & 44         & 215        \\
Validation set   & 98             & 35        & 64         & 88         & 285        \\  
\end{tabular}%
\end{table}

\subsubsection{Performance of Random Forest classifier without data balancing}
The tumour probability heatmaps for all the 500 whole slide images were generated using Ensemble-A configuration (section \ref{path_sec:ens_config}), and from the heatmaps, all the features listed in Table \ref{path_tab:metastases-features} were extracted. Post generation of features, the training set was cleaned by removing some of the outlier points. The outliers were detected based on threshold-based heuristics like the presence of significantly large tumour false-positive regions in negative cases etc.
For the purpose of classifier selection, feature elimination and hyper-parameter tuning, the classifiers were initially trained on the train set (n=215) and later validated on the held-out validation set (n=285). Experimentation on various classifiers showed that the optimal performance in terms of classification accuracy (90.18\%) and Cohen's kappa score (0.9164) on held-out validation set was achieved with Random Forest classifier with 100 trees. From Table \ref{path_tab:cm17-datasplit}, it can be observed that the data distribution was highly class imbalanced, with negative cases being the majority class and ITC cases being the minority class. This lead to misclassifications between ITC and negative cases, as evident in the confusion matrix. 

Further experimentation was performed by training another Random Forest classifier on the complete training set (n=500) in order to maximize the utilization of training points. The five-fold cross-validation showed an average accuracy score of 90\%, and its performance was similar to the model trained on train set (n=215).
The above two trained models are referred to as RF-PI and RF-CI (Random Forest classifiers trained on the partial and complete training set with imbalanced class data, respectively).


\subsubsection{Performance of Random Forest classifier after data balancing}
The train set (n=215) split and the complete training set (n=500) were separately balanced using SMOTETomek algorithm and two Random Forest classifiers were trained using these two balanced datasets. The two trained models are referred to as RF-PB and RF-CB (Random Forest classifiers trained on Partial and Complete training set, which are Balanced data, respectively). Table \ref{path_tab:RF-val_study} provides the results of the validation study performed on all four models. It was observed that post data balancing of the training dataset, the 5-fold cross-validation accuracy scores improved by a margin of 5\%.

\begin{table}
\centering
\caption{The table provides the validation results of the four Random Forest classifiers, each trained on different subsets of the training data. Note: For the models RF-PI and RF-PB, held-out validation existed, whereas, for the other two models, it was not available as it was trained on the entire training set, and hence N.A (not applicable) is mentioned in the table. For all the models, 5-fold cross-validation accuracy was estimated on their respective training sets. These values are provided as mean (standard deviation).}
\label{path_tab:RF-val_study}
\begin{tabular}{@{}lll@{}}
  
                    & \multicolumn{2}{c}{\textbf{Accuracy (\%)}}   \\    
\textbf{Classifier} & \textbf{5-fold CV} & \textbf{Validation set} \\
RF-PI               & 87 (0.06)          & 90.18                   \\
RF-PB               & 92 (0.03)          & 87.02                   \\
RF-CI               & 89.89 (0.03)       & N.A                     \\
RF-CB               & 94.83 (0.02)       & N.A                     \\  
\end{tabular}%
\end{table}


\subsection{Tumour segmentation analysis on DigestPath dataset}
 
\begin{figure}
    \includegraphics[width=0.7\columnwidth]{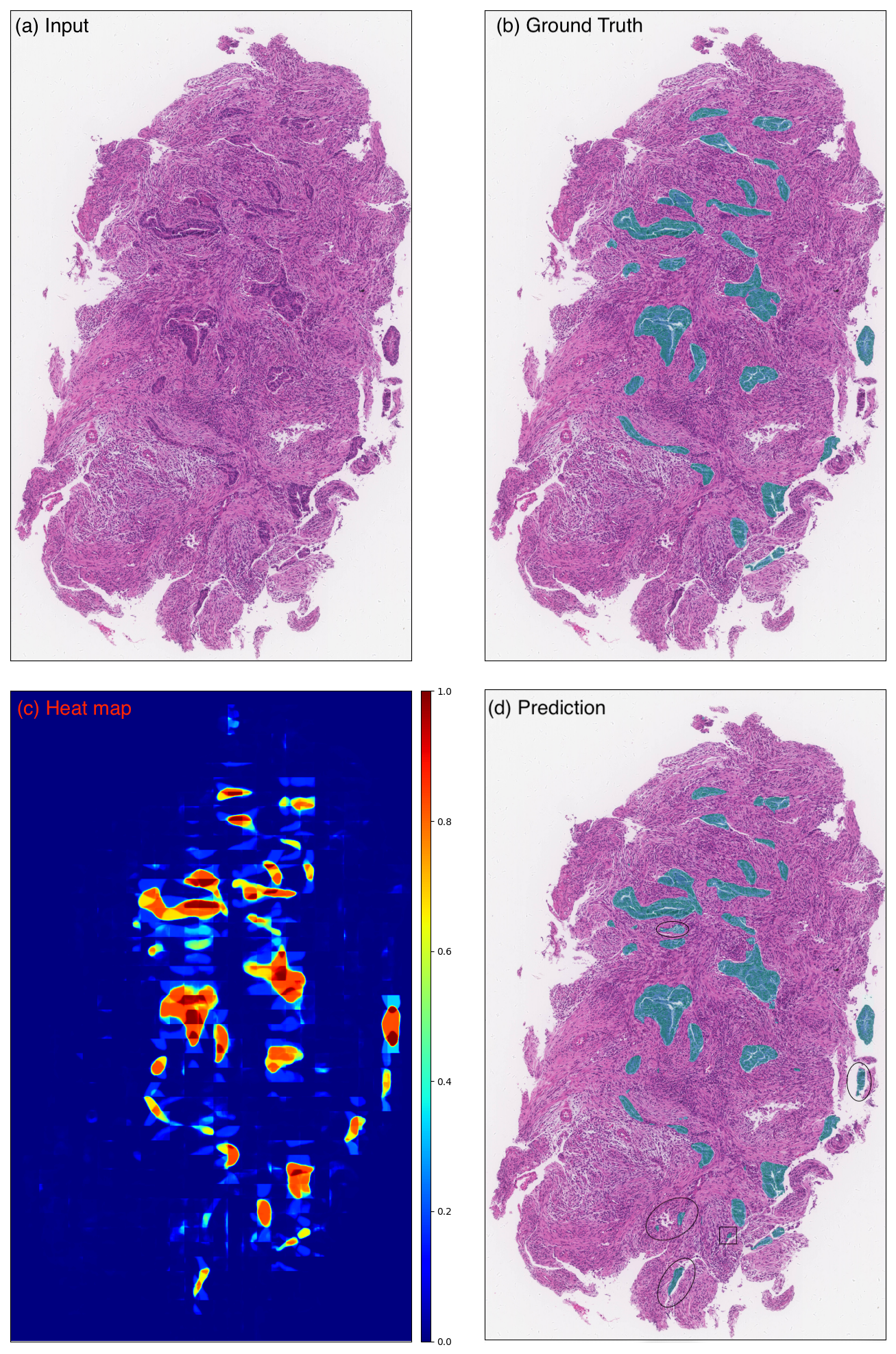}
    \centering
    \caption{An illustration of the segmentation results on a sample from DigestPath dataset. (Clockwise from top left) Whole slide image of H\&E stained colon cancer tissue; Pathologist annotated ground truth of the tumour overlayed on the whole slide image (green region indicates the tumour); Segmentation map overlayed on the whole slide image (probability map thresholded at 0.5); Heatmap of Tumour probability. The detected false positives are circled (d).}
    \label{path_fig:digestpath_hd}
\end{figure}

\begin{table}
\centering
\caption{Segmentation results on the held-out validation set (n=25) of DigestPath dataset.}
\label{path_tab:digestpath_heldout_results}
\begin{tabular}{@{}ll@{}}
  
Model                 & Dice \\    
DeepLabV3Plus          & 0.81 \\
DenseNet-121 FCN       & 0.84 \\
Inception-ResNet-V2 FCN & 0.84 \\
Ensemble               & 0.86 \\  
\end{tabular}%
\end{table}
In this section, details specific to the training and inference strategies on DigestPath dataset are presented. Out of 660 tissue images from DigestPath training set, 635 images were used for training, and the remaining images were used as held-out validation set (n=25) for model selection and hyperparameter tuning. The training set (n=635) was split further into three-fold cross-validation sets; each set of data was used to train the individual models in the ensemble. A total of 80,000 patches from the entire training data were extracted, and each model was trained on patch size of 256x256 and a batch size of 32. The model inference procedure involved extraction of patches of size 256x256 with 50\% overlap between adjacent patches in batches of 32. In order to generate the binary segmentation map, the predicted tumour probability map was thresholded at 0.5. Figure \ref{path_fig:digestpath_hd} illustrates an example of the segmentation map generated by the proposed ensemble model. The trained models were tested on held out validation set (n=25), and the corresponding results are tabulated in Table \ref{path_tab:digestpath_heldout_results}.

\subsection{Tumour segmentation analysis on PAIP dataset}

\begin{figure}
    \includegraphics[width=1.\columnwidth]{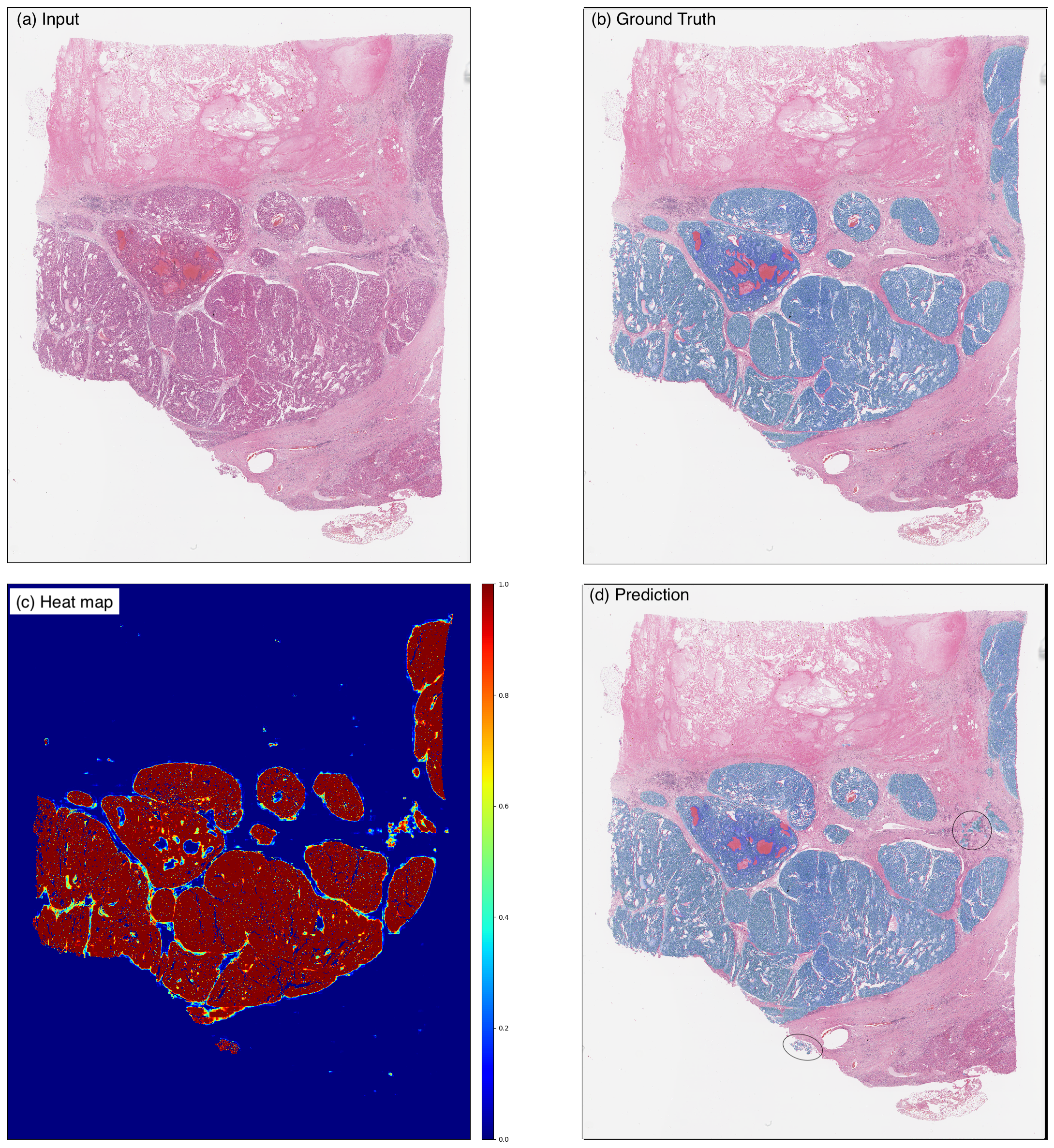}
    \caption{An illustration of the segmentation results on a sample from PAIP dataset. (Clockwise from top left) Whole slide image of H\&E stained liver cancer tissue; Pathologist annotated ground truth of the tumour overlayed on whole slide image (blue region indicates the tumour); Segmentation map overlayed on whole slide image (probability map thresholded at 0.5); Heatmap of Tumour probability. The detected false positives are circled (d).}
    \label{path_fig:paip_hd}
\end{figure}
In this section, details specific to the training and inference strategies on PAIP dataset are presented. The tissue mask generation (as detailed in \ref{path_sec:tm}) incorporated the post-processing step of closing morphological operation with a kernel size of 20, followed by an opening operation with a kernel size of 5 and a final dilation operation with a kernel size of 20. The training data set (n=50) was split into five-fold cross-validation and out of these five-folds only three of them were used for training the models of the ensemble. The data was split into five-folds as opposed to three-folds to ensure that each training set had at least 40 samples. A total of 200,000 patches from the entire training dataset were extracted with equal contributions from each training sample. The models were trained with a patch size of 256 and a batch size of 32.  The model inference procedure involved extraction of non-overlapping patches of size 1024x1024 in batches of 16. For the generation of segmentation maps, the generated tumour probability maps were thresholded at 0.5. 
The threshold was decided based on the experimental analysis for a range of threshold values on the validation set (n=10). The optimal threshold value was found to be 0.5. It was observed that lower thresholds resulted in false positives in samples and higher thresholds led to under-segmentation. Figure \ref{path_fig:paip_hd} illustrates an example of the segmentation map generated by the proposed ensemble configuration. The performance of the trained models on the validation set (n=10) released by the challenge organisers is tabulated in Table \ref{path_tab:paip_validation_results}.

\begin{table}
\caption{Segmentation results on the validation set (n=10) of PAIP dataset.}
\label{path_tab:paip_validation_results}
\centering
\begin{tabular}{@{}ll@{}}
  
Model              & Jaccard Score \\    
DeepLabV3Plus      & 0.681786      \\
Inception-ResNet-V2 FCN & 0.685771      \\
DenseNet-121 FCN           & 0.679107      \\
Ensemble           & 0.701621      \\  
\end{tabular}
\end{table}

%

\subsection{Viable tumour burden analysis on PAIP dataset}
\begin{figure}
    \centering  
    \subfloat[]{\includegraphics[width=0.9\columnwidth]{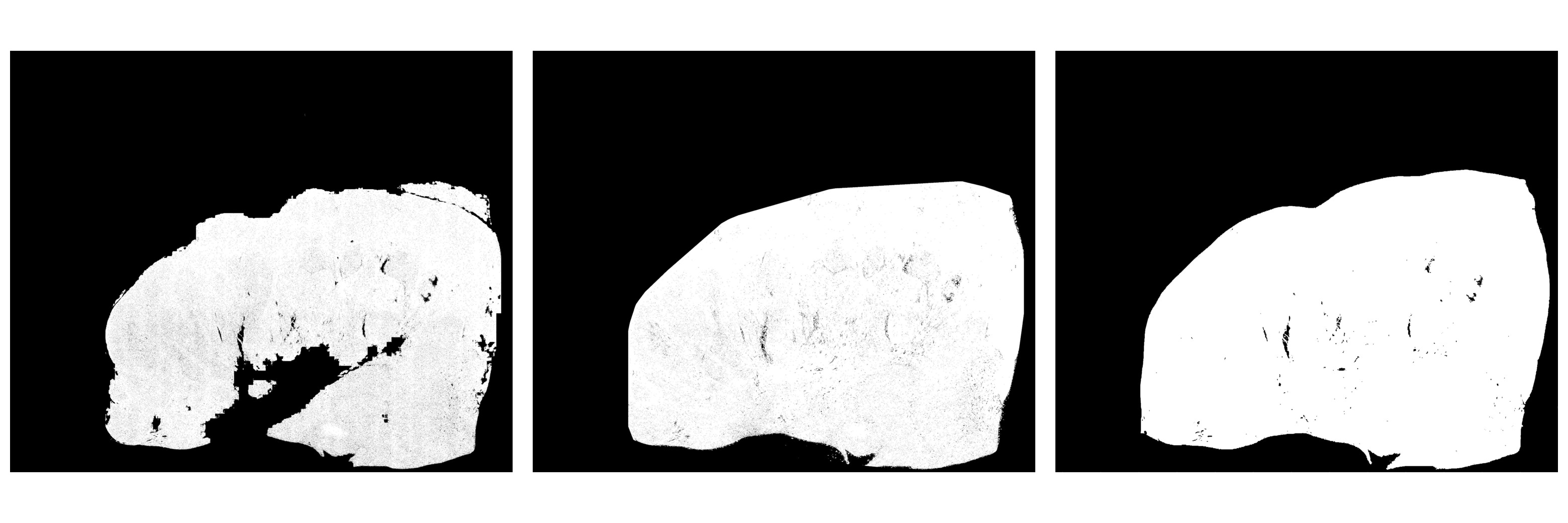}}\\
    \subfloat[]{\includegraphics[width=0.9\columnwidth]{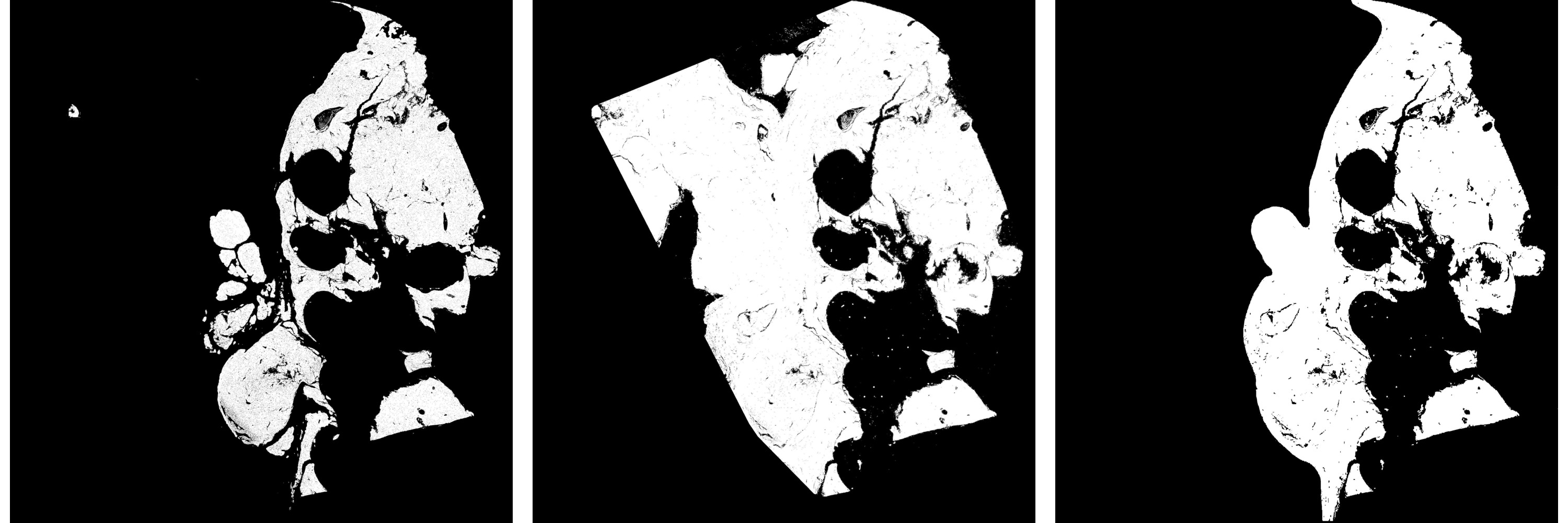}}\\
    \subfloat[]{\includegraphics[width=0.9\columnwidth]{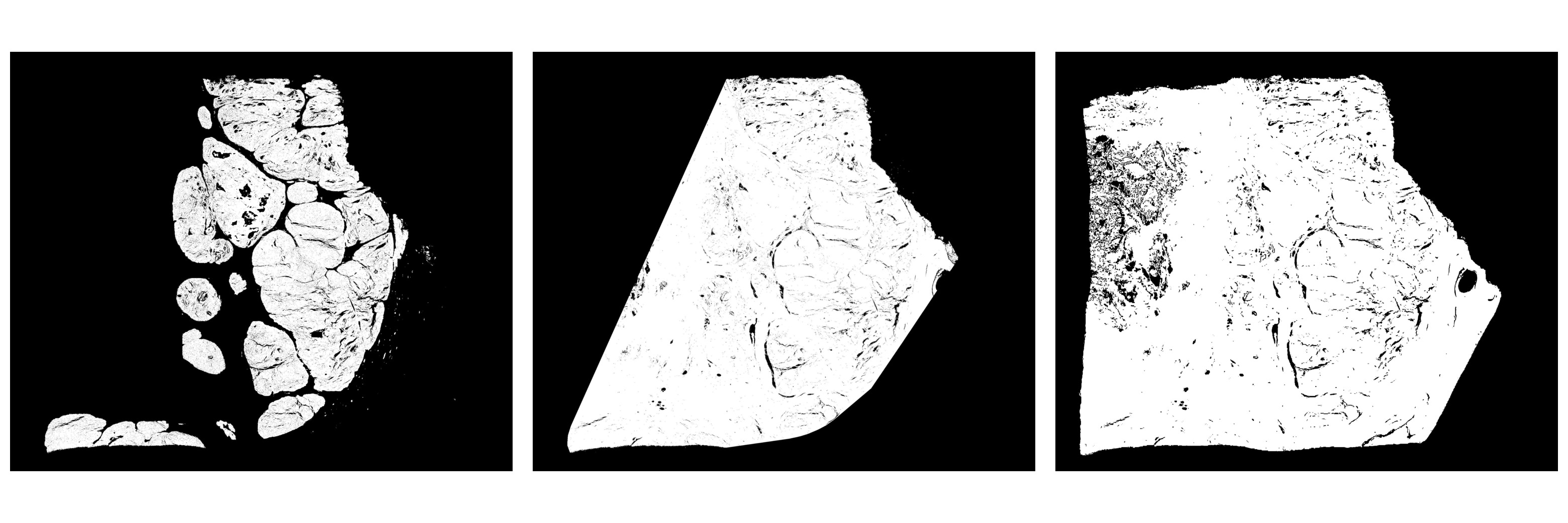}}\\
    \caption{(Left to Right) Viable tumour prediction; Whole tumour prediction; Pathologist annotated whole tumour ground truth}
    \label{path_fig:vtb-sample}
\end{figure}

Fig. \ref{path_fig:vtb-sample} shows the results obtained using the proposed methodology described in section \ref{path_sec:path_tumorburden}. In (a) the predicted whole tumour segmentation was similar to pathologist provided ground truth of whole tumour region and most of the samples in the dataset fell into this category. The proposed methodology for the whole tumour region failed in following cases where -  (b) the predicted viable tumour regions were scattered into small discrete disjoint regions which were distant from the most prominent viable tumour region and (c) the whole tumour region was larger than the convex hull of the viable tumour region.

\subsection{Uncertainty analysis}
In this section, the demonstration and interpretation of the proposed uncertainty analysis on the DigestPath dataset. Analysis of the PAIP 2019 and CAMELYON are present in the supplementary section. Fig. \ref{path_fig:camelyon_results} provides an illustration of aleatoric and epistemic uncertainty analysis on a held-out test case from the DigestPath dataset. The proposed patch-based method for aleatoric uncertainty estimated high uncertainty values inside tumour regions because of prevalent loss of neighbouring context information at patch borders; hence aleatoric uncertainty estimation necessitates the analysis to be conducted on a larger contiguous region of a whole slide image. The proposed uncertainty analysis was done with a patch size of 256 $\times$ 256 because of computational constraints. In Figure \ref{path_fig:camelyon_results} (4)(d) illustrates epistemic uncertainty maps, where the uncertain regions corresponded to the boundary surrounding the tumour tissue. In Figure \ref{path_fig:camelyon_results} (5)(d) illustrates the map of combined uncertainties (average of aleatoric uncertainties for all the three models along with epistemic uncertainty across the three models). In Figure \ref{path_fig:camelyon_results} (4,5)(c), indicates that the ensemble prediction reduced the number of false positives, thereby increasing overall Dice score to 0.94, which is about 0.04-0.06 improvement in Dice score when compared to the individual models.

\begin{figure}
    \includegraphics[width=1.\columnwidth]{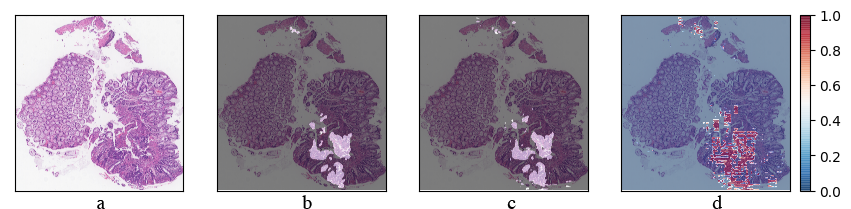}
    \caption{(a) Colon cancer tissue sample from DigestPath dataset, (b) Pathologist annotation of tumour overlayed on whole slide image, (c) Tumour probability heatmaps overlayed on the whole slide image and, (d) Aleatoric uncertainty maps.}
    \label{path_fig:digestpath_results}
\end{figure}

\subsection{CAMELYON uncertainty maps}
\begin{figure*}
\centering
    \subfloat[(1) DenseNet-121 FCN model predictions with aleatoric uncertainty (Dice = 0.91). ]{\includegraphics[width=0.8\textwidth]{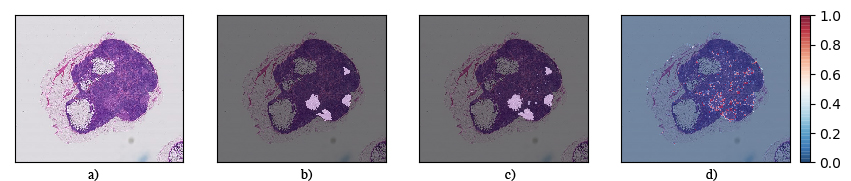}}\\
    \subfloat[(2) Inception-ResNet-V2 FCN model predictions with aleatoric uncertainty (Dice = 0.89) ]{\includegraphics[width=0.8\textwidth]{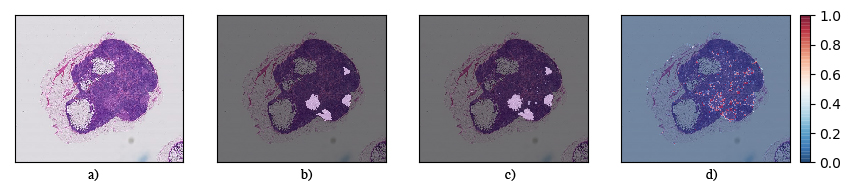}}\\
    \subfloat[(3) DeepLabv3Plus model predictions with aleatoric uncertainty (Dice = 0.88). ]{\includegraphics[width=0.8\textwidth]{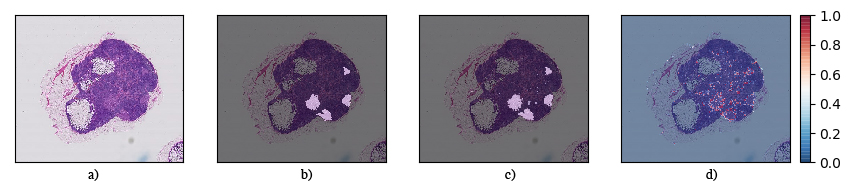}}\\
    \subfloat[(4) Ensemble model predictions with epistemic uncertainty (Dice = 0.94). ]{\includegraphics[width=0.8\textwidth]{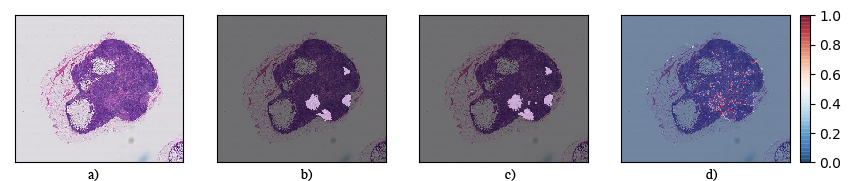}}\\
    \subfloat[(5) Ensemble model predictions with combined uncertainty (Dice = 0.94). ]{\includegraphics[width=0.8\textwidth]{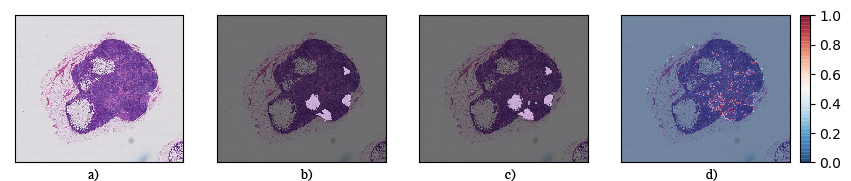}}
    \caption{In the figure for 1-5 (a) Whole slide image of H\&E stained lymph node section from CAMELYON dataset, (b) Pathologist annotation of tumour overlayed on whole slide image, (c) Tumour probability heatmaps overlayed on the whole slide image and, (d) Corresponding uncertainty analysis maps.}
    \label{path_fig:camelyon_results}
\end{figure*}

Figure \ref{path_fig:camelyon_results} provides an illustration of aleatoric and epistemic uncertainty analysis on a held-out test case from the CAMELYON dataset.

\section{Challenge results}
\label{path_sec:path_results}
\subsection{Performance evaluation on CAMELYON17 challenge}
On the CAMELYON17 testing dataset (n=500) the ensemble strategy was employed by combining the predictions from all the four trained Random Forest classifiers. The ensembling was based on the majority voting principle, and in case of a tie, the higher metastases category was selected. The ensemble model is referred to as RF-Ensemble. Table \ref{path_tab:cm17-pnstage} compares the results of the proposed ensemble approach with other published approaches on CAMELYON17 testing dataset (n=500). The proposed ensemble strategy gave Cohen's kappa score of 0.9090.

\begin{table}
\centering
\caption{Comparison of the proposed with other published approaches for automated pN-Staging in CAMELYON17 challenge. The score reported in the table is from the open public leader board of CAMELYON17 challenge. The proposed approach (RF-Ensemble) stood rank-3 on the leader-board  (Accessed on 31-Dec-2019). The table additionally provides the performance of individual Random Forest classifiers in the ensemble and RF-Ensemble classifier.}
\label{path_tab:cm17-pnstage}
\begin{tabular}{@{}lll@{}}
  
\textbf{Method}                            & \textbf{Cohen Kappa Score} & \textbf{Rank} \\    
\citep{leeautomatic}       & 0.9570                     & 1             \\
\citep{pinchaudcamelyon17} & 0.9386                     & 2             \\
\textbf{Proposed (RF-Ensemble)}                         & 0.9090                     & 3             \\
Proposed (RF-PI)                               & 0.8971                     & 12            \\
Proposed (RF-PB)                               & 0.9027                     & 9             \\
Proposed (RF-CI)                               & 0.8889                     & 18            \\
Proposed (RF-CB)                               & 0.9057                     & 6             \\  
\end{tabular}%
\end{table}

\subsection{Performance evaluation on DigestPath 2019 challenge}
Table \ref{path_tab:digestpath_scores} compares the results of the proposed with other approaches on DigestPath-2019 testing dataset (n=212). The proposed approach obtained a Dice score of 0.78 on the test set.

\begin{table}
\centering
\caption{Top four entries in DigestPath-2019 challenge.}
\begin{tabular}{@{}ll@{}}
  
\textbf{Teams}          & \textbf{Dice}  \\    
kuanguang       & 0.807 \\ 
zju\_realdoctor & 0.792 \\ 
TIA\_Lab        & 0.787 \\ 
\textbf{Proposed}  & 0.782 \\  
\end{tabular}

\label{path_tab:digestpath_scores}
\end{table}

\subsection{Performance evaluation on PAIP 2019 challenge}
Table \ref{path_tab:paip_results} compares the results of the proposed with other approaches on PAIP-2019 testing dataset (n=40). The challenge comprised of two tasks,  described as follows-
\begin{itemize}
\item Task 1: Liver cancer segmentation performance was evaluated using average Jaccard index. 
\item Task 2: Viable tumour burden estimation was evaluated as the average of products of absolute accuracy and corresponding Task 1 score (Jaccard index) for each of the cases in the test set.
\end{itemize}
For Task 1, all the participants utilized deep learning-based methods for segmentation of viable tumour, albeit with different CNN architectures. For Task 2, all the participants used deep learning-based methods for segmentation of the whole tumour. The proposed convex hull based approximation method showed comparable performance with deep learning-based methods.

\begin{table}
\centering
\caption{Top five entries of PAIP 2019. Task 1 corresponds to Viable tumour segmentation and Task 2 corresponds to Viable tumour burden estimation. Note: FNLCR: Frederick National Laboratory for Cancer Research.}
\begin{tabular}{@{}lll@{}}
  
\textbf{Team}                                              & \textbf{Task 1} & \textbf{Task 2}  \\    
FNLCR & 0.789                 & 0.752                        \\
Sichuan University                                & 0.777                 & NA                           \\
\textbf{Proposed}                                              & 0.750                 & 0.6337                       \\
Alibaba                                 & 0.672                 & 0.6199                       \\
Sejong University                                 & 0.665                 & 0.6330                       \\  
\end{tabular}
\label{path_tab:paip_results}
\end{table}

\section{Open source contribution}
\label{path_sec:open_contrib}
\begin{figure}
    \includegraphics[width=1.\columnwidth]{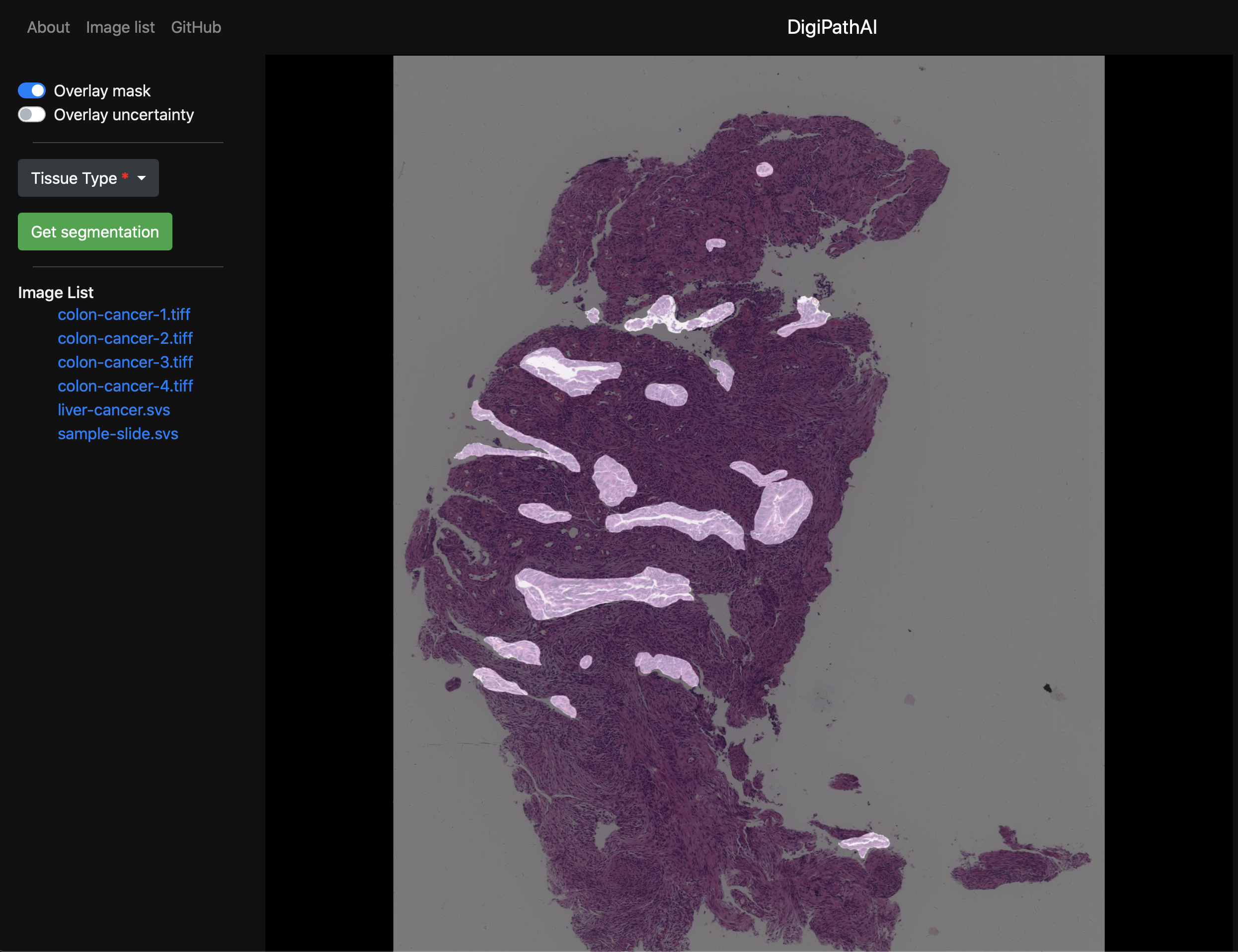}
    \caption{User interface of the whole slide image analysis software.}
    \label{path_fig:digipathai}
\end{figure}
An open-source application \citep{digipathai} on top of the proposed segmentation pipeline was developed and released. The application (Figure \ref{path_fig:digipathai}) can load whole slide images, run the segmentation algorithm, and calculate the uncertainties. The software is modular making it easy for researchers to easily add their own segmentation pipelines or extend it's functionality. 

an API with which researchers can utilize the segmentation pipeline within their applications. Conversely, the application's modular structure allows for researchers to test their segmentation pipeline with the application's GUI as well. The slide viewer was built using OpenSlide \citep{goode2013openslide} and OpenSeadragon \citep{openseadragon}.

\section{Discussion and conclusions}
\label{path_sec:discussion}
An automated end-to-end deep learning-based framework for segmentation and downstream analysis of whole slide images was developed. The proposed framework showed state-of-the-art results on three publicly available histopathology image analysis challenges, namely, CAMELYON, PAIP 2019 and DigestPath 2019. The problem of segmentation of gigapixel whole slide images was approached using the divide-and-conquer strategy by dividing the large image into computationally feasible patch sizes and running segmentation algorithms on patches and stitching segmented patches to generate the segmentation of the entire whole slide image. The patches were segmented using an ensemble of FCNs, which are encoder-decoder based architectures employed for generating dense pixel-level classification. The encoders in the proposed FCNs were some of the state-of-the-art CNNs used for natural image analysis tasks, and the decoders were a learnable upsampling module to generate dense predictions. The proposed segmentation framework was an ensemble comprising of multiple FCN architectures, each independently trained on different subsets of the training data. The ensemble generated the tumour probability map by averaging the posterior probability maps of all the FCNs. The ensemble approach showed superior segmentation performance when compared to its individual constituting FCNs. The patch-based segmentation methods for large-sized images suffer from loss of neighbouring context information at patch borders. This issue was addressed during inference by proposing- (i) to use patch size larger than that used during training and (ii) averaging overlapping patch region's posterior probability maps while stitching tumour probability maps for the entire whole slide image. In addition to the generation of tumour probability heatmaps, a methodology for generating uncertainty maps based on model and data variability was also incorporated into the framework. These uncertainty maps would assist in better interpretation by pathologists and fine-tuning the model with uncertain regions.

Further research can be done in the design of efficient and multi-resolution FCN architectures for capturing multi-resolution information from whole slide images \citep{graham2019mild}. The proposed experimental analysis on transfer learning showed that pre-training models with different histopathology datasets could act as good starting points for training models were pathology datasets are limited. Post-processing techniques could be one of the directions to improve the predicted whole slide image's tumour segmentation; techniques such as patch-based conditional random fields \citep{krahenbuhl2011efficient, li2018cancer} could be employed to refine the predicted segmentation masks rather than employing hardcoded threshold values.

The segmentation of whole slide images is usually the primary step for many analysis tasks such as metastases classification and estimation of tumour burden.  In this regard, an automated pipeline for lymph node metastases classification and pN-staging was developed. For the task of lymph node metastases classification, an ensemble of multiple Random Forest classifiers was proposed, and each classifier was trained on different subsets of the training data. The training data was prepared by extracting features based on the pathologist's viewpoint from the tumour probability maps. Additionally, incorporating synthetically generated training samples into the training data demonstrated its efficacy in addressing class imbalanced datasets for such classification tasks. 

The proposed method for viable tumour burden estimation from whole slide images of liver cancer utilized an empirical method for estimating the whole tumour region from the predicted viable tumour region. The whole tumour region was proposed to approximate a convex hull around the viable tumour region. This approximation performed on par with other deep learning-based segmentation approaches and was also computationally inexpensive. The proposed method could be refined further by incorporating learning-based methods. For example, the convex hull output could be used as an initial point for active contours-based models \citep{kass1988snakes} for correcting whole tumour region segmentation.

\bibliographystyle{model2-names}
\bibliography{refs} 

\end{document}